\documentclass[aps,pra,reprint,showpacs,noshowkeys,floatfix]{revtex4-1}

\usepackage{graphicx}
\usepackage{amsmath,amssymb}
\usepackage[backref=none]{hyperref}
\usepackage[usenames,dvipsnames]{color}
\usepackage{mathtools}

\usepackage{lipsum}

\definecolor{MyDarkBlue}{rgb}{0,0.1,0.7}
\definecolor{gray}{rgb}{0.7,0.7,0.7}
\hypersetup{pdfborder={0 0 0},colorlinks,breaklinks=true,
  urlcolor={MyDarkBlue},citecolor={MyDarkBlue},linkcolor={MyDarkBlue}
}

\newcommand{\rme}{\mathrm{e}}
\newcommand{\rmd}{\mathrm{d}}
\newcommand{\rmi}{\mathrm{i}}

\newcommand{\lamSpos}{\lambda_\Sigma^+}
\newcommand{\gamt}{{\gamma}}
\newcommand{\ncl}{{n}} 
\newcommand{\bpsi}{\boldsymbol{\psi}}

\newcommand{\eref}[1]{(\ref{#1})}
\newcommand{\fref}[1]{Fig.~\ref{#1}}

\begin{document}

\title{Genuine many-body quantum scars along unstable modes in Bose-Hubbard systems}

\newcommand{\LiegeUniversity}{${}^1$CESAM research unit, 
University of Liege, B-4000 Liège, Belgium}
\newcommand{\RegensburgUniversity}{${}^2$Institut f\"ur Theoretische Physik, 
Universit\"at Regensburg, D-93040 Regensburg, Germany}

\author{Quirin Hummel${}^{1,2}$}
\author{Klaus Richter${}^2$}
\author{Peter Schlagheck${}^1$}
\affiliation{\LiegeUniversity}
\affiliation{\RegensburgUniversity}
\date{\today}

\begin{abstract} 

The notion of many-body quantum scars is associated with special eigenstates, usually concentrated in certain parts of Hilbert space, that give rise to robust persistent oscillations in a regime that globally exhibits thermalization.
Here we extend these studies to many-body systems possessing a true classical limit characterized by a high-dimensional chaotic phase space, which are not subject to any particular dynamical constraint.
We demonstrate genuine quantum scarring of wave functions concentrated in the vicinity of unstable classical periodic mean-field modes in the paradigmatic Bose-Hubbard model.
These peculiar quantum many-body states exhibit distinct phase-space localization about those classical modes.
  Their existence is consistent with Heller’s scar criterion and appears to persist in the thermodynamic long-lattice limit.
Launching quantum wave packets along such scars leads to observable long-lasting oscillations, featuring periods that scale asymptotically with classical Lyapunov exponents, and displaying intrinsic irregularities that reflect the underlying chaotic dynamics, as opposed to regular tunnel oscillations.

\end{abstract}


\maketitle

The past decade has witnessed tremendous progress in the understanding of the key mechanisms that inhibit thermalization in complex quantum many-body systems.
While many-body localization generically arises in the presence of a significant amount of disorder and/or interaction \cite{GorMirPol05PRL,BasAleAlt06AP,OgaHus07PRB,ZniProPre08PRB,LukO19S}, nongeneric phenomena of weak ergodicity breaking, typically manifested by persistent oscillatory behaviour of observables \cite{BerO17N}, can occur in systems that globally exhibit eigenstate thermalization in the considered parameter regime.
Such ergodicity breaking behaviour is generally attributed to \textit{scarring} \cite{TurO18NPhys}, a concept that was originally introduced in single-particle chaotic quantum systems exhibiting two degrees of freedom \cite{Hel84PRL,Bog88PhD}.
A scar in the proper sense refers to a quantum eigenstate that is semiclassically anchored on an unstable periodic orbit \cite{Hel84PRL,Bog88PhD} instead of being equidistributed over the entire chaotic phase space as predicted by the eigenstate thermalization hypothesis \cite{Deu91PRA,Sre94PRE}.
As argued by Heller \cite{Hel84PRL}, such a scarred eigenstate can exist provided the period $T$ of the orbit is relatively short and its Lyapunov exponent $\lambda$ relatively weak, i.e. $T \lambda \lesssim 2\pi$, such that a wave packet that is launched along this orbit will almost recover its original shape after one period.
Scars are not to be confused with ordinary ``regular'' quantum states anchored on stable periodic orbits, whose existence and characteristics are most straightforwardly inferred from Einstein-Brillouin-Keller quantization rules \cite{BerTab76PRSA}.

\begin{figure}[t]
\includegraphics[width=\linewidth]{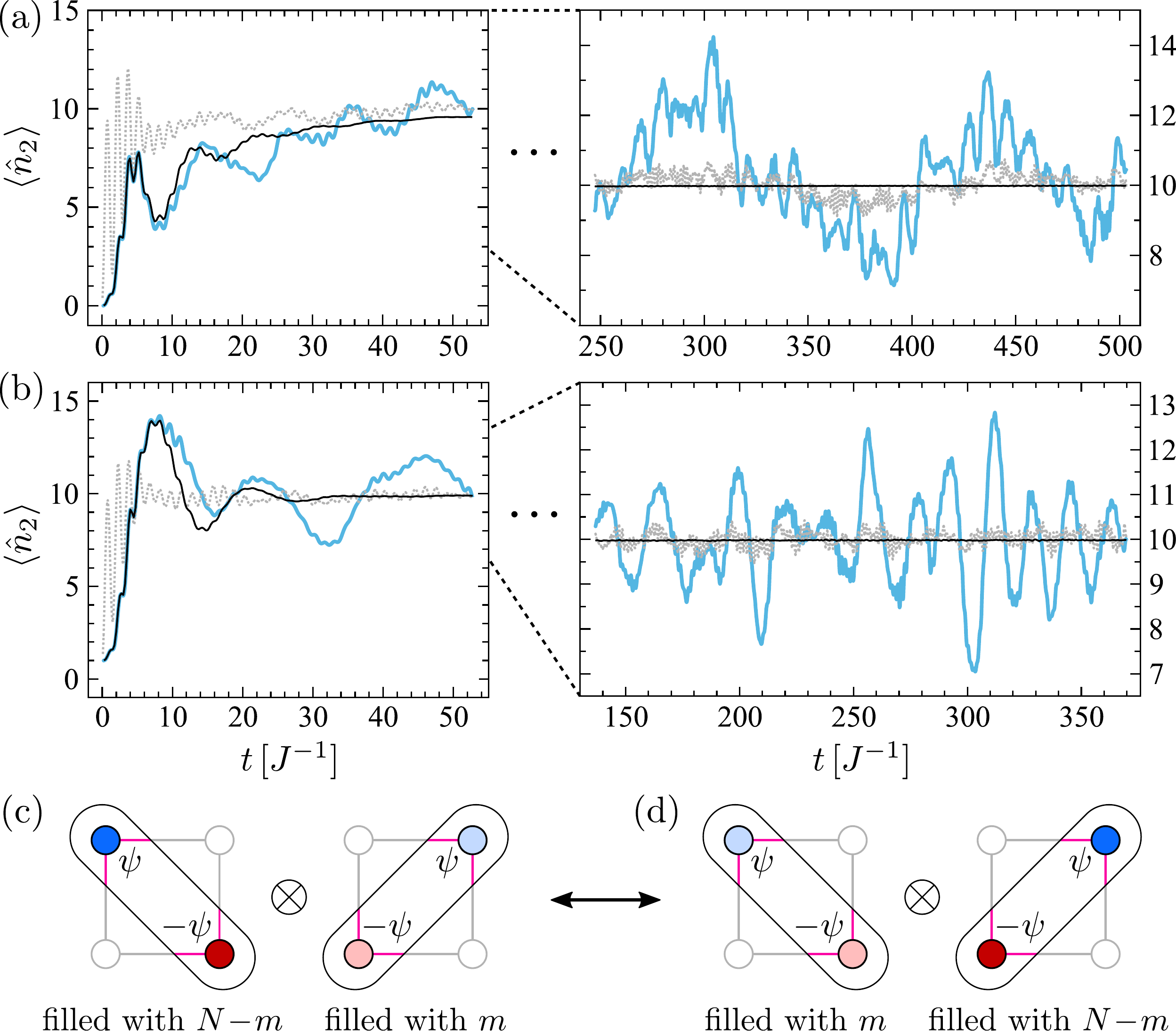}
\caption{\label{fig:osc}
  {\bf Persistent oscillations reflecting quantum chaotic scarring} --
  (a,b): Time evolution of the on-site occupancy $\langle \hat{n}_2 \rangle$.
  When initialized in staggered dimer product states $\lvert \pi_m \rangle$ [see Eq.~\eref{eq:prod}] the system oscillates [blue lines, with (a) $m=0$ and (b) $m=2$] between coupled  partners (c) $\lvert \pi_m \rangle$ and (d) $\lvert \pi_{N-m} \rangle$, symmetry-related by $90^\circ$-rotation of the plaquette.
  In contrast, classical dynamics (as implemented by TWA, black) as well as quantum evolution of initial Fock states $\lvert \frac{N-m}{2}, \frac{m}{2}, \frac{N-m}{2}, \frac{m}{2} \rangle$ (dotted gray) undergo fast thermalization (for particle number $N\!=\!40$ and effective interaction $\gamt \!=\!0.95$).}
\end{figure}

The recent discovery of many-body scars in quantum simulators \cite{BerO17N,BluO21S}, followed by numerous theoretical studies employing spin chains, $PXP$ models, or dynamically constrained systems (e.g.~\cite{MouRegBer18PRB,SchIad19PRL,CHaO20PRB,DesO21PRL,TurO18NPhys,HoO19PRL,KheLauCha19PRB,ChoO19PRL,HudO20CP,ZhaO20PRL,SuO22X,HudO22PRB}, see \cite{SerAbaPap21NPhys,Pap22,MouBerReg22RPP} for recent reviews), calls for an investigation of those scar characteristics in the high dimensional domain.
In the context of the widely considered spin-chain like systems such a study is however hampered by the fact that those quantum Hamiltonians do not have an obvious classical counterpart that would naturally arise from a semiclassical evaluation of Feynman's path integral \cite{Gutzwiller}.
Artificial classical phase spaces can nevertheless be constructed using the time-dependent variational principle \cite{HoO19PRL}, by which means unstable periodic orbits associated with many-body scars can indeed be identified. 

To extend the investigations of many-body scars to quantum systems possessing a true classical (chaotic) limit, we propose here to study many-body scars in Bose-Hubbard (BH) systems, whose  high-dimensional classical counterpart is well defined and given in terms of a discrete nonlinear Schrödinger equation.
Unlike other recent studies on scarring in BH systems \cite{HudO20CP,ZhaO20PRL,SuO22X}, we shall not consider a dynamically constrained configuration owing to the presence of correlated hopping, periodic driving, or tilting, but study unconstrained homogeneous rings of finite size, square plaquettes in particular.
The scars that we find there are anchored on unstable classical staggered dimer configurations, for which the exchange of population between adjacent lattice sites is dynamically suppressed despite nonvanishing hopping matrix elements.
As is shown in Fig.~\ref{fig:osc}, a preparation of a quantum state on such a classical configuration gives rise to persistent oscillatory behaviour in one-body observables, indicating the absence of thermalization.
Irregular features are identified in these oscillations [panels (a),(b)], in line with the high dimensionality of the underlying chaotic phase space in which the dynamics takes place. 

We consider $N$ interacting bosonic particles confined to a one-dimensional periodic lattice of $L$ wells. 
This system is described by the BH Hamiltonian
\begin{equation} \label{eq:H}
  \hat{H} = - J \sum_{l=1}^L (\hat a^\dagger_{l+1} \hat a_l + \hat a^\dagger_{l} \hat a_{l+1}) + \frac{U}{2} \sum_{l=1}^L \hat n_l ( \hat n_l - 1 )
\end{equation}
with bosonic on-site creation-, annihilation-, and number operators $\hat a^\dagger_l$, $\hat a_l$, and $\hat{n}_l = \hat a^\dagger_l \hat a_l$, where we have nearest-neighbor hopping $J$, repulsive on-site interaction $U>0$ and periodic boundary conditions, $l \in \mathbb{Z}_L$.
It formally admits a well-defined classical limit where the system is described by a condensate wave function $\bpsi = ( \psi_1 , \ldots, \psi_L ) \in \mathbb{C}^L$ whose time evolution is governed by the discrete non-linear Schr{\"o}dinger equation (DNLSE)
\begin{equation} \label{eq:DNLSE}
  \rmi \dot\psi_l = - J ( \psi_{l-1} + \psi_{l+1} ) + U (\psi^\ast_l \psi_l - 1) \psi_l
\end{equation}
(setting $\hbar\! =\! 1$).
The latter is obtained as saddle point equa\-tion in the path integral formulation of the quantum system \cite{EngO14PRL,DubMue16NJP,Richter2022}, yielding the quantum-to-classical mappings $\hat a_l \mapsto \psi_l$, $\hat a^{\dagger}_l \! \mapsto \! \psi^{\ast}_l$, and $\hat{n}_l \!+\! 1/2 \! =\! (\hat a_l\hat a^{\dagger}_l + \hat a^{\dagger}_l\hat a_l)/2 \mapsto |\psi_l|^2$.
As a purely classical description, Eq.~\eqref{eq:DNLSE} becomes formally valid in the mean-field regime of large average site occupancies $N/L\to \infty$ and small $U \to 0$, scaled such that the effective dimensionless interaction parameter
\begin{equation} \label{eq:gamt}
  \gamt = (N/L + 1/2) U / J
\end{equation}
is kept fixed.
Up to a scaling of the time $t$ and a constant shift in energy, $\gamt$ is the only parameter of the DNLSE.

We now focus on site numbers $L$ that are multiples of four.
In that case, the \textit{staggered dimer} configuration, generally characterized by a wave function of the form 
\begin{equation} \label{eq:stdim}
\bpsi = (\psi_1,\psi_2,-\psi_1,-\psi_2,\psi_1,\psi_2,-\psi_1,-\psi_2,\ldots)
\end{equation} 
for a pair of complex amplitudes $\psi_1,\psi_2$, represents a fixed point of the site occupancies in the framework of the classical DNLSE \eqref{eq:DNLSE}.
Despite a nonzero $J > 0$, hopping is dynamically suppressed in this configuration, and the site amplitudes $\psi_l$ feature only phase oscillations with frequencies $\omega_l = U (n_l - 1)$ at constant $n_l = |\psi_l(0)|^2$. 
While this resembles the Mott-insulator physics of $U/J \to \infty$, it is here a result of a fragile balance, crucially depending on the equal populations and relative phase of $\pi$ between next-nearest neighbors.
A slight deviation from the staggered-dimer manifold $\mathcal{M}_{\mathrm{SD}}$, given by all $\bpsi$ of the form \eqref{eq:stdim}, breaks this balance, leading to population transfer that may further push the system away from this manifold.
As a result, staggered dimer waves are, in a wide parameter range, at the same time fundamental short and unstable periodic modes, thus representing excellent candidates for scarring.

Semiclassically, the phenomenon of scarring is generally described as concentration of particular eigenstates of the Hamiltonian along unstable periodic orbits of the corresponding classical system that are at least locally embedded in a patch of chaotic motion.
In the present context, ``periodicity'' of a mean-field solution has to be understood modulo a global phase, i.e., we call $\bpsi(t)$ \textit{periodic} with period $T$ if for some (irrelevant) $\theta \in \mathbb{R}$
\begin{equation} \label{eq:pO}
  \bpsi( t + T ) = \bpsi( t ) \rme^{\rmi \theta} \, .
\end{equation}
To test whether a given state is scarred by such a periodic orbit $\bpsi(t)$, or, more generically, by a family of such orbits defined in a finite range of energy, we employ so-called tube states \cite{Ver12PRL,RevO12PRE} constructed as
\begin{equation} \label{eq:tubes}
  \lvert \mathcal{T}_{\bpsi(t)} \rangle \equiv \mathcal{N} \int_0^{T} \!\rmd t \, \rme^{\rmi [S( t ) - \pi \mu( t )/2] } \, \lvert \bpsi( t ) \rangle_N \, .
\end{equation}
Here we define the number-projected coherent state
\begin{equation} \label{eq:Nproj}
  \lvert \bpsi \rangle_N = \frac{1}{\sqrt{N!}}( \mathbf{e}_{\bpsi} \cdot \boldsymbol{a}^\dagger )^N  \lvert 0 \rangle \propto \hat{\Pi}_N \rme^{ \bpsi \cdot \boldsymbol{a}^\dagger } \lvert 0 \rangle
\end{equation}
centered about the phase-space point $\bpsi$, with $\mathbf{e}_{\bpsi} \!\equiv\! \bpsi /\! \sqrt{ \bpsi \cdot \bpsi^\ast }$, $\lvert 0 \rangle$ the vacuum state, and $\hat{\Pi}_N$ the projector to the $N$-particle sector.
These tube states are forced to be concentrated along the trajectory by placing a wave packet $\lvert \bpsi(t) \rangle_N$ at each of its points.
The dressing with a phase factor determined by classical dynamics, containing the accumulated classical action $S$ and Maslov index $\mu$, ensures constructive interference of neighboring wave packets and is here especially devised for oscillator-like systems
\cite{supp}.
Demanding the wave packets at $t=0$ and after time $T$ to be in phase gives the Bohr-Sommerfeld (BS) type quantization condition 
\begin{equation}
  S(T) - \pi \mu(T) / 2 + N \theta \equiv 0 \; \pmod{2 \pi} \,,
\end{equation}
singling out a discrete set of quantized orbits and corresponding tube states for each family of periodic orbits.

To ease discussions, we first focus on the simplest case $L\!=\!4$, corresponding to a single square plaquette, and thus consider the manifold of staggered dimers given by condensate wave functions of the form $\bpsi = ( \psi_1, \psi_2, -\psi_1, -\psi_2 )$.
Then the above semiclassical construction yields quantized tube states \eref{eq:tubes} with an intriguing structure.
Specifically, one finds that the $m$\nobreakdash-th quantized staggered dimer tube state $\lvert \mathcal{T}_m \rangle$, $m \in \{0, \ldots, N\}$, starting with maximal population of sites $l\!=\!1, 3$ at $m\!=\!0$, is very well described by a product state of the form
\begin{equation} \label{eq:prod}
  \lvert \mathcal{T}_{m} \rangle \simeq \lvert \pi_m \rangle \equiv \bigl\lvert (\psi_1, -\psi_1) \bigr\rangle_{N-m}^{(1,3)} \otimes \bigl\lvert (\psi_2, -\psi_2) \bigr\rangle_{m}^{(2,4)} \, .
\end{equation}
The two factors are states \eref{eq:Nproj} on the Hilbert subspaces living on sites $(1,3)$ and $(2,4)$, respectively [see Fig.~\ref{fig:osc}(c),(d)].
As a direct product of number states on disjoint subspaces, the states \eref{eq:prod} do not show any phase coherence between the two diagonals $(1,3)$ and $(2,4)$, as is classically evident from the different phase velocities $\omega_1$ and $\omega_2$, whereas the phase relation between the two opposite sites within each diagonal is fixed to $\pi$.

A characteristic hallmark for the existence of many-body scars anchored on staggered dimers can indeed be found by the numerical propagation of quantum many-body wave packets that are initialized on the states \eref{eq:prod}. 
As shown in Fig.~\ref{fig:osc}, persistent oscillations, displaying no decay over very long time scales (blue lines), arise in the mean site occupancies, in contrast to classical simulations based on the Truncated Wigner Approximation (TWA) that would predict rapid relaxation to thermal equilibrium (black).
Note that such a relaxation behaviour would also occur (grey) if the wave packet was initialized on a Fock state $\lvert \nu_1,\nu_2,\nu_1,\nu_2 \rangle$ having the same mean site occupancies $\nu_1 = (N-m)/2$ and $\nu_2 = m/2$ as the state \eref{eq:prod}.
This demonstrates the importance of the specific structure of the staggered dimer states for the occurrence of scarring.

\begin{figure}[t]
\includegraphics[width=\linewidth]{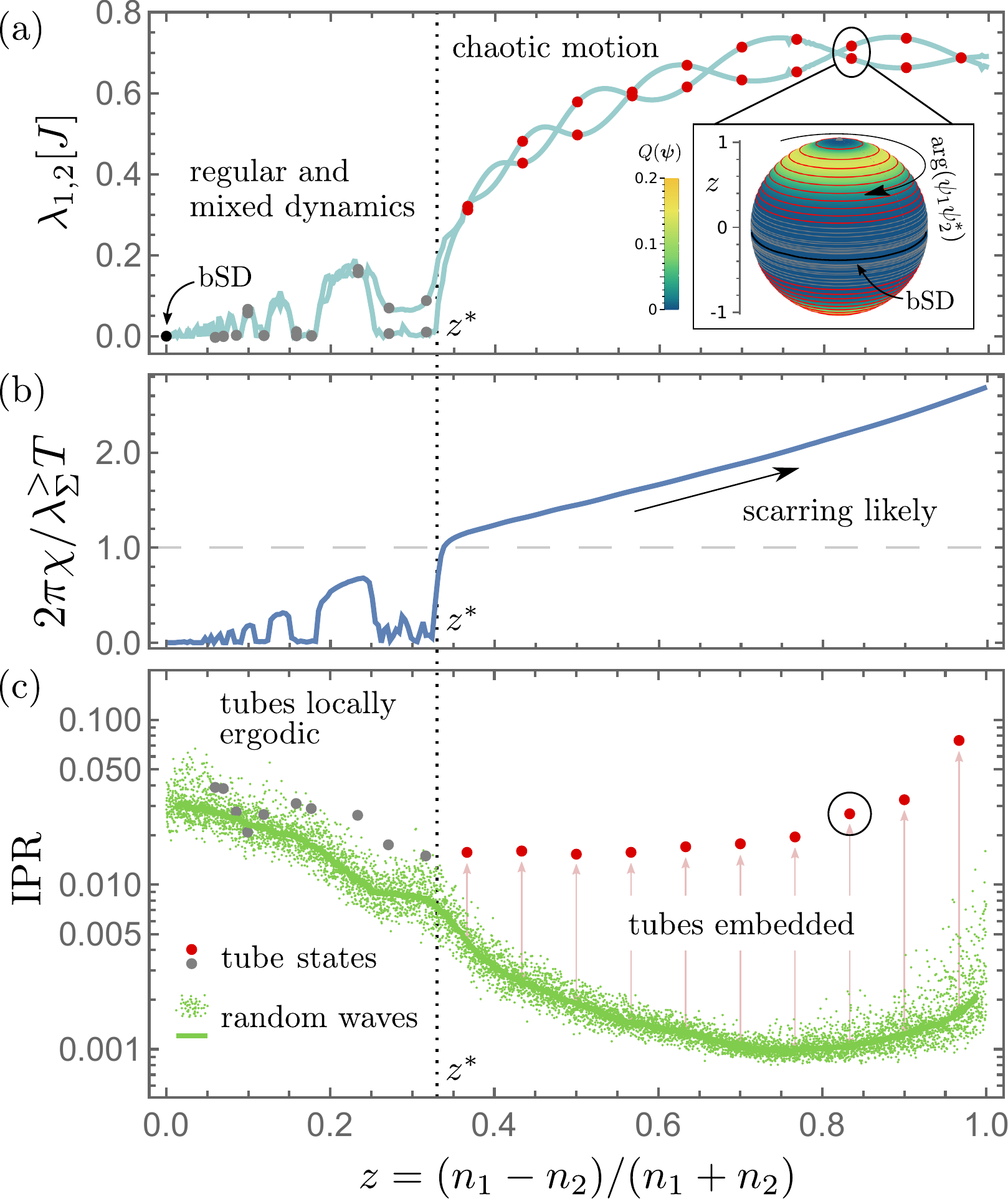}
\caption{\label{fig:zVS}
  {\bf Indicators for scarring} --
  (a) Dominant stability exponents, $\lambda_{1,2} \geq 0$, along the staggered-dimer manifold.
  Central orbits of Bohr-Sommerfeld (BS)-quantized tubes are marked by dots.
  Inset: Husimi section of one quantized tube $\mathcal{T}_2^-$ in the manifold mapped to a Bloch sphere. 
  (b) Heuristic rating for the likelihood of quantum scarring based on the Heller criterion (see text).
  (c) Inverse participation ratio \eref{eq:IPR} of BS-quantized tube states (black dots) and random-wave states (green dots) on local constant-energy layers (see text).
  As a guide to the eye, a running median is added (solid green).
}
\end{figure}

Further confirmation for the existence of genuine quantum scarring on staggered dimers is obtained via several (semi-)classical indicators, which are evaluated in \fref{fig:zVS} as functions of the imbalance $z = (\ncl_1 - \ncl_2) / (\ncl_1 + \ncl_2)$.
For the chosen intermediate coupling $\gamt = 0.95$, Eq.~(\ref{eq:gamt}), we find that quantum scars are likely to occur, independently of $N$, for imbalances $z \gtrsim z^\ast$ with $z^\ast \simeq 0.33$ (dotted vertical line), where dynamics is chaotic as indicated by \fref{fig:zVS}a).
The likelihood for scarring increases when approaching the maximally imbalanced limit due to ever shorter periods $T$.
This is demonstrated in \fref{fig:zVS}b), where we use an a-priori indicator $2 \pi \chi / \lamSpos T > 1$ for periodic orbits to support quantum scars, generalizing the heuristic Heller criterion \cite{Hel84PRL} for two-dimensional single-particle systems \cite{supp,HumSch22JPA}.
Here, $\lamSpos \equiv \Sigma_j^> \lambda_j$ is the sum of positive stability exponents, and $\chi \equiv \prod^>_j 2\lambda_j/(\lambda_\mathrm{th} + \lambda_j)$ is a heuristic factor to suppress close-to-regular or mixed dynamics with a threshold chosen as $\lambda_\mathrm{th} = 0.3 J$.

Additionally, we investigate in \fref{fig:zVS}c) the phase-space localization of the corresponding tube states by means of the inverse participation ratio (IPR)
\begin{equation} \label{eq:IPR}
  {\rm IPR}_{\lvert \phi \rangle} \equiv \mathcal{N}	\int \!\rmd^{2L}\!\psi \, \delta\!\bigl( ||\bpsi||^2 - L/2 - N \bigr) Q_{\lvert \phi \rangle}(\bpsi)^2
\end{equation}
of a state $\lvert \phi \rangle$, defined in terms of the Husimi function $Q_{\lvert \phi \rangle}( \bpsi ) \equiv \bigl\lvert \langle \phi \vert \bpsi \rangle_N \bigr\rvert^2$, which is independent of the single-particle basis owing to canonical invariance.
We compare the IPR of tube states to the one of random-wave states spread across the local layers of constant energy in which the corresponding relevant orbits are embedded.
These locally ergodic states are obtained similar to the tubes \eref{eq:tubes} but with slightly perturbed initial conditions, putting wave packets $\lvert \bpsi(t) \rangle_N$ with random phases after finite time steps $\rmd t \mapsto \Delta t$ and following the classical dynamics long enough to saturate the local constant-energy surface.
While in the regime $z \lesssim z^\ast$ of mixed regular and chaotic dynamics the tubes are found to fill up thin constant-energy layers alike random waves, they are, as seen in Fig.~\ref{fig:zVS}(c), significantly more localized than the latter in the dominantly chaotic regime $z > z^\ast$, thereby confirming their nonclassical scar-like nature
\footnote{In principle, minor chaotic layers in the mixed regime, e.g., for $0.2\lesssim z \lesssim 0.25$ in \fref{fig:zVS}a,b, may also be able to host quantum scars for sufficiently large total particle number $N$.
For all investigated parameters, however, a clear characterization of tube states as strongly embedded within locally ergodic patches by means of Eq.~\eref{eq:IPR} was only possible for the steadily chaotic regime $z \gtrsim z^\ast$.}.

\begin{figure}[t]
\includegraphics[width=\linewidth]{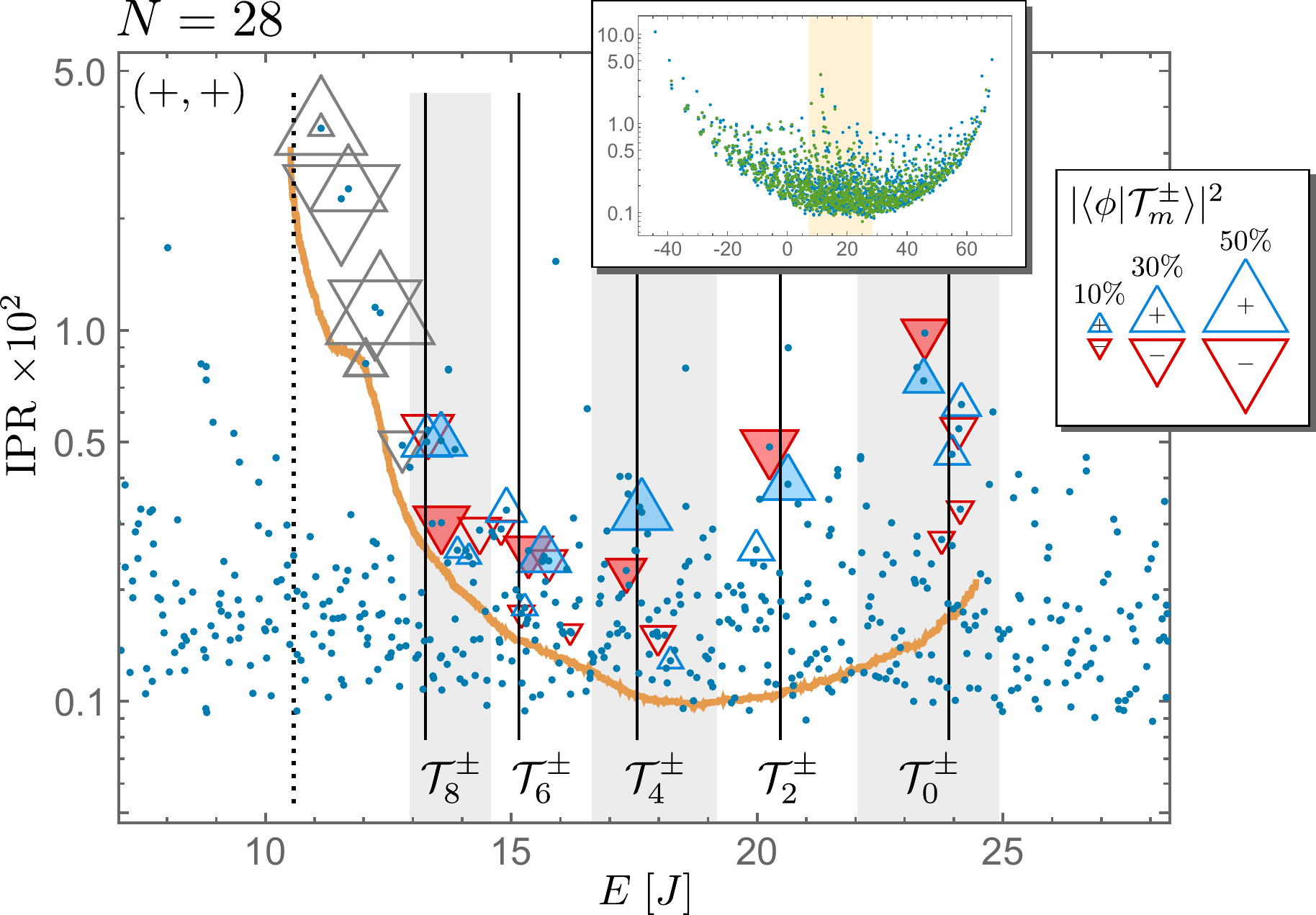}
\caption{\label{fig:IPR}
  {\bf Phase-space localization of eigenstates} -- as measured by the IPR \eref{eq:IPR} for $\gamt=0.95$ and $N=28$ in the central energy range where staggered dimer waves are located [highlighted regions in the full-range spectra $(+,+)$ and $(-,-)$ shown as inset].
  The overlap $| \langle \phi \vert  \mathcal{T}^\pm_m \rangle |^2$ of eigenstates $\lvert \phi \rangle$ with the symmetric and anti-symmetric tube states, whenever greater than 0.1, is indicated by the size of triangles pointing up and down, respectively.
  In the unstable regime $z \gtrsim z^\ast$, these are shown in blue and red, respectively, with shading and labels marking eigenstates of dominant overlap.
  Locally ergodic random waves about $\mathcal{M}_{\mathrm{SD}}$ (see \fref{fig:zVS}c) are shown in orange.}
\end{figure}

To confirm the existence of actual quantum scarring of staggered dimer solutions, we perform exact diagonalization and examine all individual eigenstates for their phase-space localization and overlap with tube states.
Figure \ref{fig:IPR} shows again the case $\gamt=0.95$, for which the energy range relevant for staggered dimers lies in the central spectrum of highly excited states, as indicated by the highlighted region of the inset.
We focus on the subspectra with the symmetry $(+,+)$, denoting fully even parity with respect to the two diagonal exchange operations (sites $1 \leftrightarrow 3$ as well as $2 \leftrightarrow 4$) of the plaquette.
For even particle number $N$ this symmetry class is shared by the staggered dimer tube states of \emph{even} BS quantization index $m$ [odd $m$ gives tube states of odd parity $(-,-)$].
We find a small number of eigenstates that are anomalously localized as compared to the majority of eigenstates with comparable energy.
As confirmed by strong overlaps with tube states, a big part of these can be directly identified as a class of staggered-dimer-like states.
Apart from some genuinely ``regular'' states featuring a high IPR due to localization on classically stable phase-space structures which are, hence, not scars according to the definition of this concept \cite{Hel84PRL}), this class also contains eigenstates that are strongly concentrated along \emph{unstable} staggered-dimer solutions embedded in \emph{chaotic} portions of the phase space.
Since their number scales proportionally to $N$ as does the number of tube states~\eref{eq:prod}, they constitute a vanishing fraction of the full spectra with Hilbert-space dimension $\sim N^{L-1}$ as $N\! \to \!\infty$.
We thus find all criteria for genuine quantum scarring fulfilled.
Note that even in the deep quantum regime of very few particles, where quantum-to-classical correspondence is no longer expected to hold, we can unambiguously identify \emph{direct descendants} of genuine quantum scars by maintaining the link between tube states and eigenstates while successively lowering $N$ \cite{supp}.

Let us discuss the absence of regularity in the oscillatory behaviour of the mean site occupancies, shown in \fref{fig:osc}.
They oscillate due to the (anti\nobreakdash-)symmetry of eigenstates with respect to the rotation of the lattice by one site, induced by the operator $\hat{R}_1$, such that scar states in the Hamiltonian's eigenspectrum exhibit a strong overlap with the two linear combinations $\lvert \mathcal{T}^{\pm}_m \rangle \propto (1 \pm \hat{R}_1) \lvert \mathcal{T}_m \rangle$.
Hence, the preparation of the quantum system on a staggered-dimer state with broken symmetry, such as $\lvert \mathcal{T}_m \rangle$, is expected to give rise to Rabi-like oscillations between $\lvert \mathcal{T}_m \rangle$ and $\hat{R}_1 \lvert \mathcal{T}_m \rangle$, with a frequency that corresponds to the level splitting of the two eigenstates $\lvert \mathcal{T}^{\pm}_m \rangle$.
This simplified reasoning is to be amended due to the fact that several eigenstates can generally be scarred with the same orbit \cite{Hel84PRL,RevO12PRE}. An initial product state $\lvert \pi_m \rangle$ gives thus rise to a superposition of corresponding frequencies and amplitudes, resulting in beatings that do not feature a clean harmonic behaviour.
In a semiclassical picture, the beating period can be estimated to be related to the classical rate to leave (or approach) the vicinity of one of these orbits, i.e., to be proportional to their inverse stability exponents $\lambda_j$ \cite{HumO19PRL}.
We confirm this scaling for the regime of weak to moderate interactions \cite{supp} where all classical stability exponents tend to be equal to a unique Lyapunov exponent, $\lambda_j \simeq \lambda_\mathrm{L}$, such that a uniform time scale $\sim \lambda_\mathrm{L}^{-1}$ emerges.
Most notably, this oscillatory behaviour is in stark contrast to the more pronounced and regular oscillations that one finds in a regime of locally \emph{stable} or close-to-stable classical dynamics, where by definition scars cannot occur \cite{supp}.

Scarring on the staggered dimer configuration \eqref{eq:stdim} is not restricted to the 4-site plaquette but can be found also for larger systems \cite{supp}, such as the $L\!=\!8$ site BH ring, where scars are anchored on wave functions of the form $\lvert (\psi_1, -\psi_1,\psi_1, -\psi_1) \rangle_{N-m}^{(1,3,5,7)} \otimes \lvert (\psi_2, -\psi_2, \psi_2, -\psi_2) \rangle_{m}^{(2,4,6,8)}$, as well as the $L\!=\!12$ site ring.
In both cases, similar irregular long-period oscillations are encountered as for the $4$-site plaquette \cite{supp}.
The staggered-dimer modes in those high-$L$ BH rings are found to have very similar Lyapunov exponents $\lambda_j \sim \gamt J$ and periods $T \sim \pi / \gamt J$, yielding a $\gamma$-independent Heller-type indicator $2 \pi / \lamSpos T \sim 2 / (L-2)$ that scales inversely with the number of chaotic degrees of freedom transverse to the mode.
This would \textit{a priori} predict a decreasing likelihood for the existence of staggered-dimer scars with increasing $L$. However, we expect this effect to be counterbalanced by the increasing number $\nu = L/4$ of discrete rotational symmetries that staggered dimer configurations feature.
The associated quantum states live in the corresponding symmetry subspaces whose dimensions are consequently lowered by a factor $\propto 1/L$ with respect to the full Hilbert space and which thus exhibit a reduced density of states as compared to the latter.
This reduction factor is expected \cite{supp} to effectively enhance the otherwise deficient Heller-type indicator to a sufficient extent, yielding support for the existence of staggered dimer scars in the thermodynamic long-lattice limit; see, e.g., scarring within a $1.35\times10^6$-dimensional Hilbert space in the specific case of a $L=12$ site BH ring \cite{supp}.

In summary, we present solid evidence for the existence of genuine scars in a preeminent bosonic many-body system that is not subject to any dynamical constraint, namely a homogeneous disorder-free BH ring. 
These scars form in the vicinity of the classical staggered dimer configuration \eqref{eq:stdim} where population exchange between sites is dynamically suppressed despite a nonvanishing hopping parameter.
The time evolution of quantum states launched on such staggered dimers reveals an intriguing feature that we conjecture to be generic for many-body scars in a high-dimensional chaotic phase space, namely the existence of persistent long-period oscillations that do not exhibit a well identifiable regularity.
This feature is open to experimental verification within state-of-the-art quantum simulators employing ultracold bosonic atoms in optical lattices \cite{LukO19S}. There, staggered-dimer product states \eqref{eq:prod} can be created by quantum quenches starting, e.g., from spatially separated left- and right-diagonal sublattices that are brought together at $t=0$ to form the plaquette.
We believe that scarring is a generic phenomenon in high-dimensional bosonic many-body systems exhibiting chaotic dynamics, and our study lays proper foundations for their unambiguous identification and characterization.

We are grateful for inspiring discussions with Steven Tomsovic, Denis Ullmo, and Juan Diego Urbina.
This research was supported by the University of Li\`ege under Special Funds for Research,
IPD-STEMA Programme, and by the Deutsche For\-schungs\-ge\-mein\-schaft through project Ri681/15-1 within the Reinhart-Koselleck Programme.

\newpage

\section*{Supplemental material}

\appendix

\renewcommand{\thefigure}{\Alph{section}\arabic{figure}}
\newcommand{\sectionQ}[1]{\section{#1} \setcounter{figure}{0}}

\newcommand{\psep}{\,}

\newcommand{\vek}{\mathbf}
\newcommand{\rmT}{\mathrm{T}}

\newcommand{\lamL}{\lambda_{\mathrm{L}}}
\newcommand{\heff}{\hbar_{\mathrm{eff}}}
\newcommand{\gsym}{g_{\mathrm{sym}}}

\newcommand{\tE}{t_\mathrm{E}}
\newcommand{\Mstab}{\mathbb{M}}

\newcommand{\Leref}[1]{Eq.~(\ref{#1}) of the main text}
\newcommand{\Lfref}[1]{Fig.~\ref{#1} of the main text}

\sectionQ{Details of the tube state construction}
\label{app:tubes}
The construction of tube states as auxiliary quantum states that are concentrated along a given periodic orbit of the underlying classical system, governed by the discrete non-linear Schr\"odinger equation (DNLSE), relies solely on classical properties of that orbit.
A commonly used prescription \cite{suppRevO12PRE} is to place isotropic Gaussian wave packets along the periodic trajectory, dressed with a phase factor that accounts for the accumulation of the classical action $S(t)$ and a Maslov-type phase $\mu(t)$.
After one full period, $t=T$, these have to coincide with the \textit{canonical} action $S(T) = \oint \rmd {\bf q} \cdot {\bf p}$ and the Maslov phase $\mu(T)$ \cite{suppRevO12PRE} in order to guarantee that the trajectory fulfills the correct Bohr-Sommerfeld quantization condition if and only if the wave packet after one full period is in phase with the initial one.
While $S(T)$ and $\mu(T)$ are thus uniquely determined canonical invariants, the evolution $S(t)$ and $\mu(t)$ for intermediate times $t \notin T \mathbb{Z}$ are usually chosen to fit the specifics of the system at hand.

Here, we employ a variant devised for oscillator-like systems by using the \emph{radial} action
\begin{equation} \label{app:eq:St}
    S(t) = \frac{1}{2} \int_0^t ( {\bf p} \cdot \rmd {\bf q} - {\bf q} \cdot \rmd {\bf p} )
\end{equation}
and
\begin{equation} \label{app:eq:mut}
    \mu(t) = \frac{1}{\pi} \int_0^t \rmd \arg \det ( A_t - \rmi B_t ) \,,
\end{equation}
where the matrices $A_t$ and $B_t$ are uniquely determined by the polar decomposition
\begin{equation}
    \Mstab_t = U_t \cdot P_t \equiv
        \left( \begin{array}{cc}
            A_t & -B_t \\
            B_t & A_t
        \end{array}
        \right) \cdot P_t
\end{equation}
of the Jacobian $\Mstab_t$.
The latter encodes the evolution of deviations $\delta{\bf z}(t) = ( \delta{\bf q}(t), \delta{\bf p}(t) )$ about the given trajectory ${\bf z}(t) = ( {\bf q}(t), {\bf p}(t) )$ in linearized approximation:
\begin{equation} \label{app:eq:deltaz}
    \delta{\bf z}(t) = \Mstab_t \cdot \delta{\bf z}(0) \,.
\end{equation}
While the symplectic and orthogonal matrix $U_t$ describes the winding of invariant manifolds around the reference trajectory, $P_t = ( \Mstab_t^\rmT \Mstab_t )^{1/2}$ is positive definite with eigenvalues $ \rme^{\lambda_j(t) t} $, $\lambda_j \in \mathbb{R}$.
Since $\Mstab_t$ is symplectic, its singular values come in pairs of mutual reciprocals, i.e., for each $j$ there is a $k \neq j$ with $\lambda_{k}(t) = - \lambda_j(t)$.
For a full period $t=T$, we omit the argument, writing $\lambda_j \equiv \lambda_j(T)$, which we refer to as the \emph{stability exponents} of the periodic orbit under consideration. Due to periodicity, they are identical to the Lyapunov spectrum of the orbit.

The choice \eref{app:eq:St} for $S(t)$ and \eref{app:eq:mut} for $\mu(t)$ distributes the accumulated phase most evenly along an oscillator orbit.
Under this prescription, the orbits of a simple harmonic oscillator [or its square, as in the single-site case of the DNLSE, see Eq. \eref{eq:DNLSE} of the main text] lead to Bohr-Sommerfeld quantized tube states that coincide with the exact eigenstates of the corresponding quantum Hamiltonian.

\sectionQ{Generalization of Heller's scar criterion}
\subsection{Multiple dimensions}
\label{app:sec:HCmultidim}

Here we define and generalize a heuristic criterion for the occurrence of quantum scarring on a particular unstable periodic orbit of a classical Hamiltonian system with an arbitrary number of degrees of freedom $d\geq 2$.
We closely follow Heller's argument for $d=2$ \cite{suppHel84PRL}, which is based on wave packet propagation being essentially classical for times well below the Ehrenfest time $\tE = \lamL^{-1} \log \hbar_{\mathrm{eff}}^{-1}$. Here $\lamL$ is the classical Lyapunov exponent describing the degree of instability and $\heff$ is Planck's constant divided by a typical classical action of the system.
A Gaussian wave packet that is initially localized on a point of an unstable periodic trajectory undergoes, under classical evolution for one full period $T$, simultaneously stretching and compression in the directions corresponding to the unstable and stable manifolds, respectively.
As a consequence, its quantum probability to return is a periodically recurring signal with an exponential damping $\sim \rme^{-\lambda t}$ arising from the deformation of the wave packet up to times of the order of $\tE$.

From a semi-quantitative Fourrier analysis of such a characteristic signal, Heller found an estimate for the upper bound of the number of eigenstates participating in the process in terms of the period $T$ and the damping rate $\lambda$ \cite{suppHel84PRL}.
If this approximate bound is well below the participation ratio expected from the eigenstate thermalization hypothesis \cite{suppDeu91PRA,suppSre94PRE}, 
there has to be a number of scarred eigenstates that are anomalously localized along the classical trajectory.
In order to translate the resulting heuristic criterion $\lambda T \lesssim 2 \pi$ to an arbitrary number of degrees of freedom $d \geq 2$, it is thus sufficient to determine the exponential damping of the return signal in terms of the stability matrix $\Mstab \equiv \Mstab_T$ (see Appendix \ref{app:tubes}), which encodes the deformation of the classically evolved wave packet.

To this end, we linearize the classical dynamics about the periodic orbit within a Truncated Wigner Approximation \cite{suppHumSch22JPA}, i.e., we express the classical return probability after one period as the phase-space integral
\begin{equation}
	P_{\mathrm{ret}}^{\mathrm{cl}}(T) \sim \int \rmd^{2d} x \; \exp \bigl[ - (\vek{x} - \vek{x}_0 )^\rmT ( 1 + \Mstab^\rmT \Mstab ) (\vek{x} - \vek{x}_0 ) \bigr]
\end{equation}
for a wave packet with centroid $\vek{x}_0$ and uncertainties equally distributed among position and momentum coordinates. 
Correspondingly, the classical return probability after $n$ periods is, up to a constant prefactor,
\begin{equation}
	P_{\mathrm{ret}}^{\mathrm{cl}}(nT) \sim  \det \bigl( 1 + ( \Mstab^\rmT )^n \Mstab^n \bigr)^{-1/2} \, .
	\label{eq:Pret}
\end{equation}
To proceed with our derivation, we assume for simplicity that the eigenvectors of $\Mstab$ are orthogonal, i.e., that $\Mstab^\rmT = \Mstab$.
Further, since $\Mstab$ is a symplectic matrix we can write its $2d$ eigenvalues as $\mu_j \equiv \rme^{\lambda_j T}$ with stability exponents ordered such that
$\lambda_j \geq 0$ and
$\lambda_{j+d} = - \lambda_{j}$ for $j=1,\ldots,d$.
We denote the number of independent constants of motion by $c$.
Each of them will contribute two vanishing exponents which we associate with the indices $j=d-c+1,\ldots,d$ and $j=2d-c+1,\ldots,2d$. This yields, after appropriate ordering,
\begin{equation}
	\det \bigl( 1 + ( \Mstab^\rmT )^n \Mstab^n \bigr) = 2^{2c} \prod_{j=1}^{d-c}
		( 1 + \underbrace{\rme^{ 2n \lambda_j T}}_{|\cdot| > 1} )
		( 1 + \underbrace{\rme^{-2n \lambda_j T}}_{|\cdot| < 1} ) \, .
		\label{eq:det}
\end{equation}
To extract the damping in the regime where $P_\mathrm{ret}^{\mathrm{cl}}$ decreases exponentially with $n$, we neglect $\rme^{-2n \lambda_j T}$ compared to unity. Combining
Eqs.~(\ref{eq:Pret}) and (\ref{eq:det}) then yields
\begin{equation}
P_\mathrm{ret}^{\mathrm{cl}}( nT ) \sim \rme^{- \lamSpos n T }
\end{equation}
with the sum of positive stability exponents 
\begin{equation}
\lamSpos \equiv \sum_{j=1}^{d-c} \lambda_j \equiv \sum_j^> \lambda_j \, .
\end{equation}
The generalization to $d$ dimensions of the heuristic {\em a-priori} Heller-type criterion for quantum scarring of a particular unstable periodic orbit is thus 
\begin{equation}
\label{eq:Heller}
    \lamSpos T \lesssim 2\pi \, .
\end{equation}
We use this equation to estimate the likelihood for scarring in BH models with $L=4\nu$ sites.

\subsection{Discrete symmetries}
\label{app:sec:HCsym}

    The original derivation~\cite{suppHel84PRL} of Heller's criterion and along with it the multidimensional generalization (see appendix \ref{app:sec:HCmultidim}) does not account for the presence of discrete symmetries.
    Consider a discrete symmetry in form of a finite symmetry group $G$ with a group action on the Hilbert space $\mathcal{H}$ given as the unitary transformations $\hat{U}(g)$, $g \in G$ that leave the system invariant, i.e., $\hat{U}^{\dagger} \hat{H} \hat{U} = \hat{H}$, and which correspond to classical point transformations $\mathbf{q} \mapsto \mathbf{f}_g(\mathbf{q})$ acting on the classical coordinates $\mathbf{q}$ such that $\langle \mathbf{f}_g(\mathbf{q}) \rvert \hat{U}(g) \lvert \phi \rangle = \langle \mathbf{q} \vert \phi \rangle$ for arbitrary quantum states $\lvert \phi \rangle$ and coordinate eigenstates $\lvert \mathbf{q} \rangle$.
    The Hamiltonian then generically assumes a block-diagonal structure with blocks corresponding to the independent symmetry subspaces $\mathcal{H}_{\alpha,i}$ of Hilbert space.
    These correspond to the (components of the) $N_{\mathrm{ir}}$ irreducible representations $M^{(\alpha)}$ of $G$, henceforth called \emph{irreps}.
    Each irrep $\alpha$ has a dimension $s_\alpha$ and its components $i = 1, \ldots, s_\alpha$ constitute independent Hilbert subspaces with degenerate eigenvalues of $\hat{H}$.
    Formally, there exists a basis $\{ \lvert \alpha, i, m \rangle \}$ with
    \begin{equation}
        \begin{split}
            \hat{U}(g) \lvert \alpha, i, m \rangle &= 
                \sum_{j=1}^{s_\alpha} M^{(\alpha)}_{ij}(g) \lvert \alpha, i, m \rangle \,, \\
            \hat{H} \lvert \alpha, i, m \rangle &= E^{(\alpha)}_m \lvert \alpha, i, m \rangle \,,
        \end{split}
    \end{equation}
    with $i,j$ indexing the $s_\alpha$ components of irrep $\alpha$ and $m$ indexing the different eigenstates within one component $i$ of $\alpha$.
    The total number of symmetry subspaces $\mathcal{H}_{\alpha,i}$ is $N_{\mathrm{s}} = \sum_\alpha s_\alpha$.

    Under this premise, we give a modified version of Heller's derivation, closely following the original, indicating that under certain assumptions scarring within symmetry subspaces can occur even if the Heller criterion is not fulfilled.
    For this purpose, we may jump over details that are not touched by the symmetry aspects.
    If an initial wave packet $\lvert \psi \rangle$ is centered on a point of a classical periodic orbit which lies within a certain classical symmetry subspace, it generically also inherits certain symmetry properties as a Hilbert space element at quantum level.
    In general, it can be written as a unique superposition of states belonging to the symmetry subspaces $\mathcal{H}_{\alpha,i}$.
    In other words, we have
    \begin{equation}
        \lvert \psi \rangle \in \bigoplus_{\mathclap{(\alpha,i)\in A_\psi}} \mathcal{H}_{\alpha,i} \equiv \mathcal{H}_\psi\,,
    \end{equation}
    where $A_\psi$ is the set of all symmetry subspaces occupied by $\lvert \psi \rangle$, indexed by $(\alpha,i)$.
    The central quantities of interest are the spectral intensities
    \begin{equation}
        I_n = \lvert \langle n \vert \psi \rangle \rvert^2
    \end{equation}
    of the wave packet, where $\lvert n \rangle$ are the eigenstates of the Hamiltonian.
    The basic reasoning behind Heller's criterion is to give two different estimates for these intensities: one based on Berry's conjecture, i.e., assuming the eigenstate thermalization hypothesis (ETH), and the other based on classical wave packet dynamics, which is valid for short enough times.
    When the two contradict each other, it indicates a violation of ETH, i.e., quantum scarring occurs.
   
    One way to express the ETH is to say that the phase space density of an eigenstate $\lvert n \rangle$, i.e., the Wigner transform of the density matrix $\lvert n \rangle \langle n \rvert$ basically follows the microcanonical density
    \begin{equation} \label{app:eq:rhom}
        \rho_\mathrm{m}(\mathbf{q}, \mathbf{p}) = \delta[ E - H(\mathbf{q}, \mathbf{p}) ] / D(E) \,,
    \end{equation}
    where $E$ is the energy $E_n$ of the eigenstate $\lvert n \rangle$, $H$ the classical Hamiltonian, or more precisely, the Wigner symbol of the quantum Hamiltonian, and $D(E)$ is the (total) density of states.
     
    With discrete symmetries present, only spectral intensities of states belonging to $\mathcal{H}_\psi$, $\lvert n \rangle \in \mathcal{H}_\psi$, are non-zero, which we also write as $n \in \mathcal{N}_\psi$, i.e., $\mathcal{N}_\psi$ is the set of all quantum numbers $n$ belonging to the states in $\mathcal{H}_\psi$.
    Thus we look at spectral intesities $I_n$ of eigenstates in $\mathcal{H}_\psi$, i.e., that belong to those symmetry classes $(\alpha,i) \in A_\psi$ that are occupied by the wave packet $\lvert \psi \rangle$.
    As a simplification, we will use the unmodified microcanonical density~\eref{app:eq:rhom} also for eigenstates of a specific symmetry class.
    We neglect the effect of enhanced or lowered density very close to the various symmetry manifolds in phase space which comes with each specific symmetry class $(\alpha,i)$, and which becomes arbitrarily narrow in the semiclassical limit $\hbar \to 0$.
    Under this assumption the normalization in~\eref{app:eq:rhom} with the \emph{total} density of states $D(E)$ is still consistent.
    We must not use here the reduced density of states $D_{\alpha,i}(E)$ of the symmetry subspace $\mathcal{H}_{\alpha,i}$, as can be easily seen from the normalization condition $\mathrm{tr} (\lvert n \rangle \langle n \rvert) = 1$, written as phase space integral
    \begin{equation}
        (2 \pi \hbar)^{-L} \int \rmd^L q \, \rmd^L p \, \delta[ E - H(\mathbf{q}, \mathbf{p}) ] / D(E) = 1 \,.
    \end{equation}
    As the ETH-based estimate for the non-zero spectral intensities we get thus
    \begin{equation}
        I_{n}^{\mathrm{ETH}} \simeq \left\{ \begin{array}{lll}
             S_\psi(E_n) / D(E_n) \;\;&\text{for}\;&\lvert n \rangle \in \mathcal{H}_\psi \,, \\
             0\;\;&\text{for}\;&\lvert n \rangle \notin \mathcal{H}_\psi \,,
        \end{array}
        \right.
    \end{equation}
    where $S_\psi(E)$ is the normalized energy probability distribution for the wave packet $\lvert \psi \rangle$.
    
    We turn now to the finite-resolution spectrum 
    \begin{equation}
        \epsilon_\tau(\omega) = \frac{1}{\pi} \sum_{n \in \mathcal{N}\psi} \frac{\sin[(E_n / \hbar - \omega )\tau]}{E_n / \hbar - \omega} \lvert \langle n \lvert \psi \rangle \rvert^2 \,,
    \end{equation}
    which equals the Fourier transform
    \begin{equation}
        \epsilon_\tau(\omega) = \frac{1}{2\pi} \int_{-\tau}^\tau \rmd t \, \rme^{\rmi \omega t} \langle \psi \vert \psi(t) \rangle
    \end{equation}
    of the return signal $\langle \psi \vert \psi(t) \rangle$ with resolution $\tau$.
    From classical wave packet propagation one knows that the return signal
    is a series of exponentially damped recurrences, with a time interval given by the classical period $T_\mathrm{cl}$ and the damping $\exp( - \lamSpos t / 2 )$ given by the sum $\lamSpos$ of positive stability exponents of the classical periodic orbit that the wave packet has been placed on initially.
    This is a valid description for times well below the Ehrenfest time and is not affected by symmetries.
    From this one infers that the finite-resolution spectrum, calculated for a value of $\tau$ that exceeds the time scale when the recurrences have exhausted, is a series of broadened bands, separated by the spacing $\Delta \omega = \omega_\mathrm{cl} = 2 \pi / T_\mathrm{cl}$ and each with a bandwidth $\lamSpos$.
    Recognizing that the low-resolution spectrum $\epsilon_\tau(\omega)$, with $\tau$ in between the first decay time scale and the first recurrence time scale, is equal to the normalized energy probability distribution $S_\psi(E)$ with $E = \hbar \omega$, one can estimate the intensities of eigenstates within one band.
    The total intensity $I_\mathrm{tot}^{\mathrm{b}}$ within one band (located at energy $E_\mathrm{b}$) is equal to the energy probability distribution integrated over one band-spacing,
    \begin{equation}
        I_\mathrm{tot}^{\mathrm{b}} = \hbar \Delta\omega \, S_\psi( E_\mathrm{b} ) \,.
    \end{equation}
    This total intensity has to be shared among the symmetry classes $(\alpha,i) \in A_\psi$ relevant for our wave packet.
    Assuming equal sharing among all relevant eigenstates gives the intensity in the band contributed by one symmetry class $(\alpha,i)$ as
    \begin{equation}
        I^\mathrm{b}_{\alpha,i} = \frac{ N^\mathrm{b}_{\alpha,i} } {N^\mathrm{b}_{\psi} } I^\mathrm{b}_{\mathrm{tot}} \,,
    \end{equation}
    where
    \begin{equation}
        N^\mathrm{b}_{\alpha,i} = D_{\alpha,i}(E_\mathrm{b}) \hbar \lamSpos
    \end{equation}
    is the number of eigenstates in the band that are of symmetry $(\alpha,i)$, with $D_{\alpha,i}(E)$ the density of states of the corresponding subspace, and where
    \begin{equation}
        N^\mathrm{b}_{\psi} = \sum_{\mathclap{(\alpha,i) \in A_\psi}} N^\mathrm{b}_{\alpha,i}
    \end{equation}
    is the number of eigenstates in the band that are relevant for $\lvert \psi \rangle$ as regards symmetry.
    The average intensity $I^\mathrm{b}_n = I^\mathrm{b}_{\alpha,i} / N^\mathrm{b}_{\alpha,i} = I^\mathrm{b}_{\mathrm{tot}} / N^\mathrm{b}_{\psi}$ of one (symmetry-relevant) eigenstate $\lvert n \rangle \in \mathcal{H}_\psi$ in the band is thus
    \begin{equation}
        I^\mathrm{b}_n = \frac{2\pi}{\lamSpos T} \times \frac{S_\psi(E_\mathrm{b})}{\sum_{{(\alpha,i) \in A_\psi}} D_{\alpha,i}(E_\mathrm{b})} \,.
    \end{equation}
    The ratio between the average intensity based on classical wave propagation and spectral band analysis to the estimate from the ETH is thus
    \begin{equation} \label{app:eq:HellerSym}
        \frac{ I^\mathrm{b}_n }{ I^\mathrm{ETH}_n } \simeq \frac{2\pi}{\lamSpos T} \times \frac{D(E_\mathrm{b})}{\sum_{{(\alpha,i) \in A_\psi}} D_{\alpha,i}(E_\mathrm{b})} \equiv \frac{2\pi}{\lamSpos T} \gsym \,,
    \end{equation}
    where the first term is the Heller-type indicator without symmetry (but generalized to multiple dimensions). This Heller-type indicator is then effectively enhanced due to symmetry by the factor $\gsym$ given by the ratio of the total density of states to the combined density of states of only the relevant symmetries.
    In full analogy with the conventional Heller criterion, a value of this indicator that exceeds unity would here indicate the (weak) breaking of ergodicity, i.e., that there are some eigenstates that have increased overlap with wave packets placed along the periodic orbit and exhibit an enhanced probability density along the orbit, i.e., they violate the eigenstate thermalization hypothesis.
    
    This finding is in line with the fact that conventional quantum scarring in two-dimensional systems has been observed to be most prominent for scars along orbits in symmetry planes~\cite{suppHel84PRL}.

\sectionQ{Scars in the deep quantum regime}

\begin{figure}
  \centering
  \includegraphics[width=0.98\linewidth]{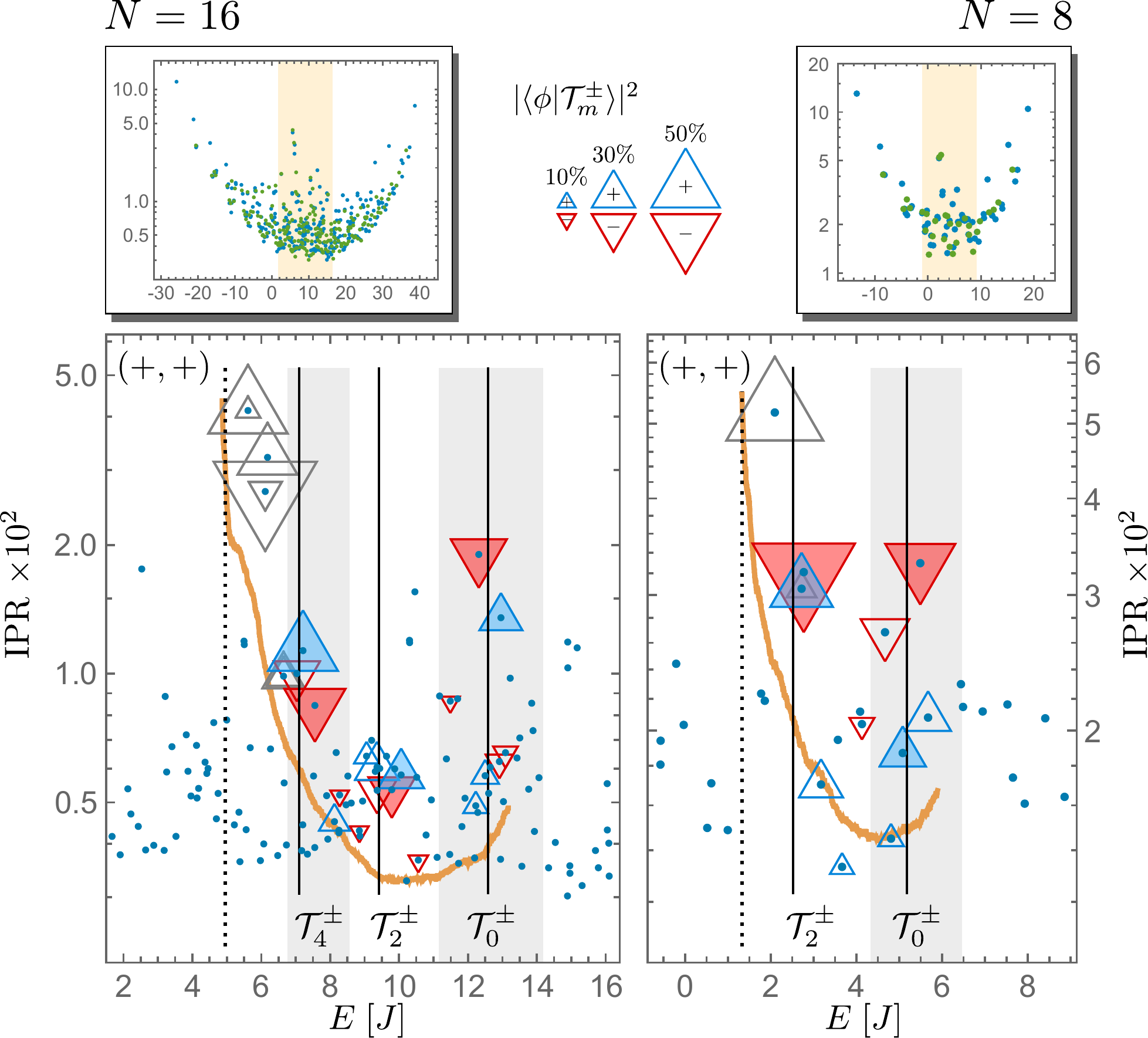}
  \caption{\label{app:fig:IPRlowN}
    Inverse participation ratio (IPR) measuring phase space localization.
    Same as \Lfref{fig:IPR} but for $N=16$ (left column) and $N=8$ (right column). 
    We can clearly identify scars anchored about the staggered dimer configuration that are descendants of the scars found for $N=28$.
  }
\end{figure}

As pointed out in the main text, scarring around staggered dimer configurations is predicted to occur in a chaotic parameter regime featuring an intermediate coupling parameter
$\gamma \!\simeq\! 1$
as well as a sufficiently strong population imbalance $z$ between the even and odd sites of the plaquette.
Keeping those two parameters fixed, we can now use the freedom of lowering the average site occupancy to 
explore how far the concept of staggered-dimer scars can be extrapolated from the semiclassical into the deep quantum regime.

In analogy with the analysis in the main text for $N/L=7$ (see \Lfref{fig:IPR}), we also find for lower average site occupancies classes of eigenstates that
exhibit strong overlap with the BS-quantized tube states anchored at staggered-dimer orbits.
In a certain regime, these classical solutions are again unstable, such that the criteria for quantum scarring would technically be fulfilled.
However, when leaving the semiclassical regime one has to be careful here and keep in mind that in the deep quantum regime the discrimination of quantum states by localization in phase space---and in particular along periodic orbits---can become inadequate due to the small Hilbert space dimension.
 Consequently, an eigenstate could have a strong overlap with a tube state of an unstable orbit out of pure coincidence, whereas the corresponding analysis in the semiclassical limit could reveal that the two features are not causally related.
 
\begin{figure*}
  \centering
  \includegraphics[width=0.7\linewidth]{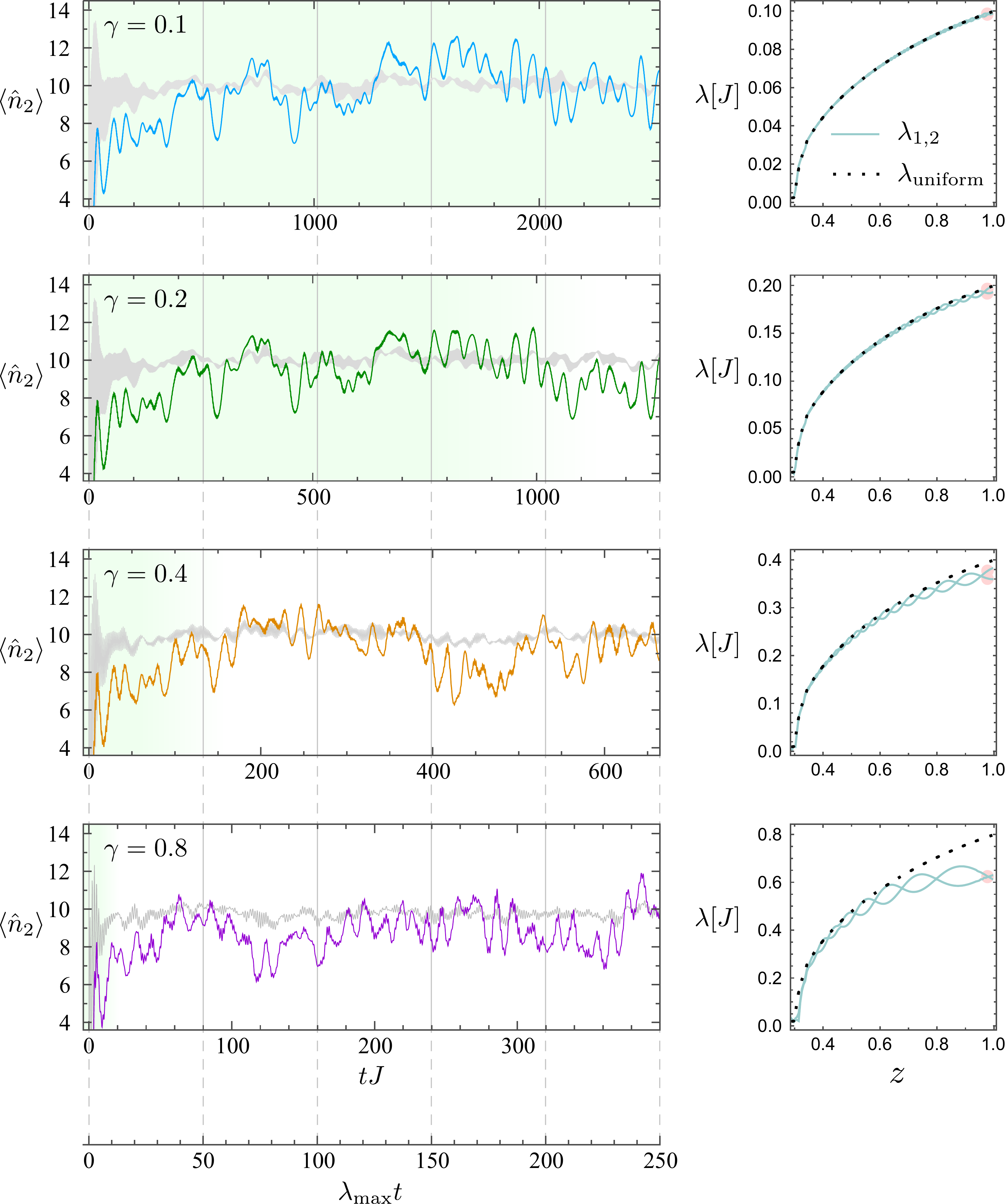}
  \caption{\label{app:fig:occs_uniform}
    Dynamics of the expectation value of a single-site occupancy in the Bose-Hubbard plaquette with $N=40$ atoms and four different interaction strengths  $\gamt=0.1, 0.2, 0.4$, and $0.8$ (left panels, from top to bottom).
    When prepared in staggered dimer product states $\lvert \pi_m \rangle$ (here $m=0$, i.e., $z=1$, colored blue, green, orange, and purple), the persistent oscillations attributed to quantum scarring tend towards a universal non-trivial function $n^\infty_l$ of $t \lambda_{\mathrm{max}}$, the time scaled with the largest stability exponent, when one approaches the regime of very weak interaction $\gamt \lesssim 0.2$ (upper panels).
    For stronger interaction, the persistent oscillations remain (in contrast to Fock states, shown in gray), although not following the universal function any longer.
    The time intervals of universality are indicated by green shading.
    This finding is explained by the classical (in)stability spectra of the staggered dimer orbits (right panels), which tend towards uniformity, $\lambda_1 = \lambda_2 \sim \gamt$, (dotted) for $\gamt \to 0$ as well as their classical period scaling accordingly (see text).
  }
\end{figure*}

In order to rule out that the found staggered-dimer signatures of eigenstates are coincidental, we decrease the number of atoms $N$ in a successive step-by-step manner and are thus able to maintain the connection between specific eigenstates and unstable staggered-dimer orbits.
Figure \ref{app:fig:IPRlowN} shows two steps with $N/L=4$ and $N/L=2$ of such a protocol performed in the plaquette with otherwise the same parameters as used in \Lfref{fig:IPR}.
We can clearly identify scar states that have very similar characteristics to the ones found for $N=28$ (see \Lfref{fig:IPR}).
Most notably, this allows us to identify eigenstates that are closer to the ``quantum many-body scar'' phenomenology as direct descendants of genuine quantum scars in the many-body system with increasing atom number.

\begin{figure*}
  \centering
  \includegraphics[width=0.8\linewidth]{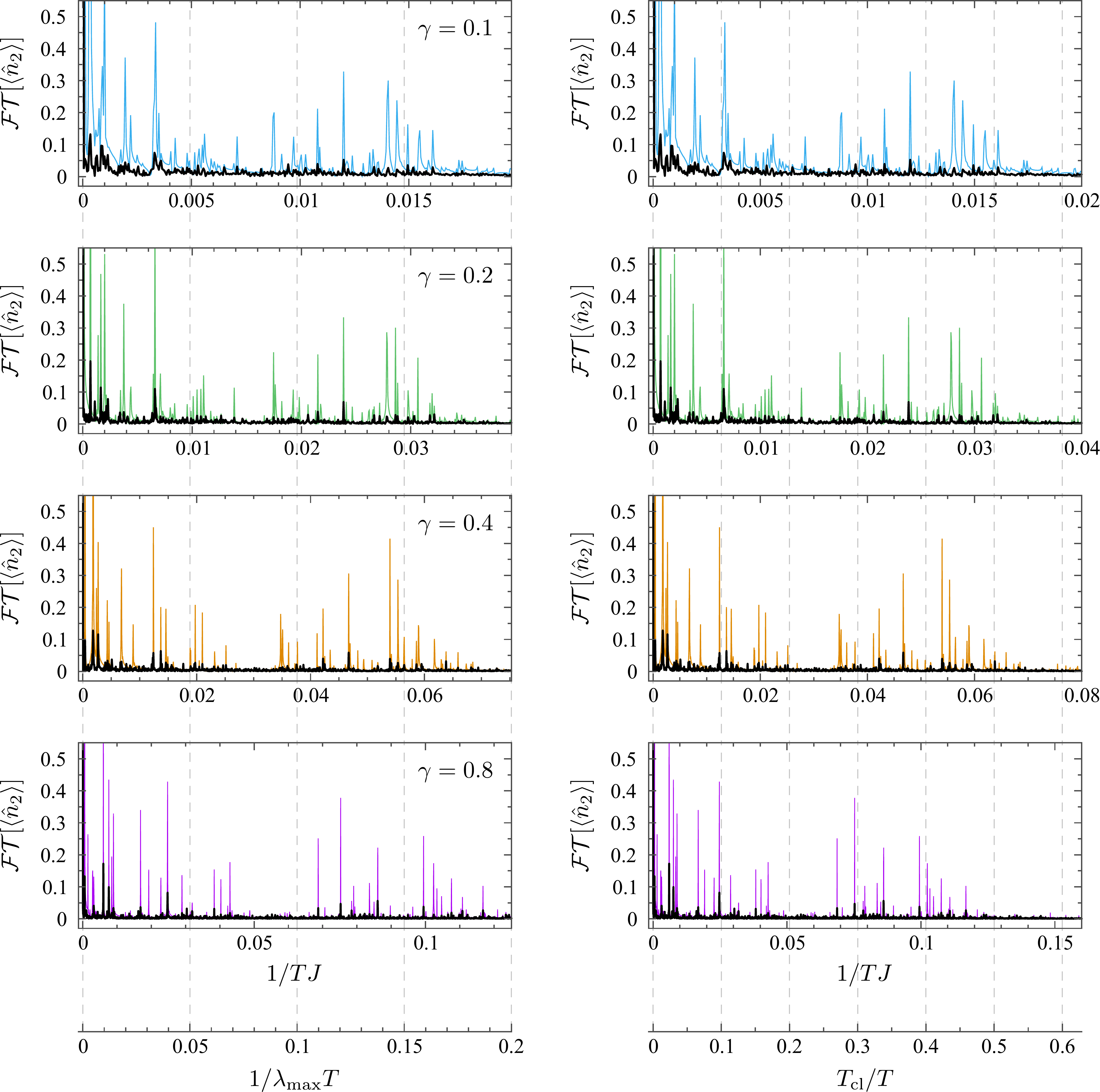}
  \caption{\label{app:fig:occs_uniform_FT}
    Fourier spectra of long time occupancy oscillations with parameters as in \fref{app:fig:occs_uniform}.
    The frequency windows are scaled with $\lambda_{\mathrm{max}}$ shown on the left panels.
    The right panels show the same spectra with frequencies $1/T$ scaled with the inverse classical staggered-dimer period $1/T_{\mathrm{cl}}$ instead.
    For very weak interactions $\gamt \lesssim 0.2$, all peaks scale almost identically (upper four panels).
    For stronger interactions, scaling with $\lambda_{\mathrm{max}}$ and $1/T_{\mathrm{cl}}$ yields different spectra and the peak frequencies apparently scale in a more individual fashion (see text).
  }
\end{figure*}

\sectionQ{Uniform time scale for weak interactions}
\label{app:sec:uniform}

In the weakly interacting regime $\gamt\lesssim 0.5$ we find that a uniform time scale emerges that governs the dynamics of imbalanced staggered dimers.
As shown in \fref{app:fig:occs_uniform}, changing $\gamt$ in the occupation dynamics of the Bose-Hubbard plaquette initialized in the product state $m=0$ results in a scaling of time. For very weak interaction the
non-trivial form of the signal due to superimposed processes of various frequencies remains unchanged in a broad time range.
The weaker the interaction, the longer it takes until residual dephasing leads to deviations from a universal limiting function
$n^\infty_l( \tau )$:
\begin{equation}
    \langle \hat{n}_l( t ) \rangle \simeq n^\infty_l( t / t_0( \gamt ) ) \psep,
\end{equation}
valid up to some break time $t \lesssim t_{\rm break}^{\infty}( \gamt )$ with $t_{\rm break}^{\infty} / t_0 \to \infty$ as $\gamt \to 0$.
This scaling behaviour is strongest for coupling parameters $\gamt \lesssim 0.2$. There the correspondence with a unique wave form $n^\infty_l$ is clearly visible up to long times $\lambda_{\mathrm{max}} t > 200$ 
with maximum Lyapunov exponent $\lambda_{\mathrm{max}}$ (indicated by green shading within panels a and b of \fref{app:fig:occs_uniform}).
For $\gamt = 0.4$ (panel c), the correspondence still holds for shorter times
$t_{\rm break}^{\infty} \simeq 100$--$200$.
For interactions as strong as $\gamt = 0.8$ (panel d), the breaking time becomes comparable to the dominant periods in the beatings, such that the scaling behavior cannot be observed anymore.
Note, however, that the responsible dephasing happens only between a relatively small set of discrete frequencies.
It therefore does not lead to a damping of the oscillatory behaviour, i.e., thermalization, as it generically occurs when initializing the system in ``typical'', non-scarred states, such as Fock states in the site basis (shown in grey).

We attribute this behaviour of universal time scaling to a peculiarity of the underlying classical system.
For small $\gamt$, we find that the stability spectrum of positive Lyapunov exponents of the staggered dimer orbits becomes uniform, $\lambda_1(\gamt) \to \lambda_2(\gamt)$ (see right panels in \fref{app:fig:occs_uniform}).
Furthermore, the unique limiting stability exponent scales linearly with the coupling strength, $\lamL( \gamt ) \sim \gamt$ as $\gamt \to 0$.
At the same time, the period of the staggered dimer orbits scales inversely proportional, $T_{\rm cl} = \pi / \gamt z J$.
As a consequence, all available classical time scales become inversely proportional to $\gamt$.

To further corroborate this, in \fref{app:fig:occs_uniform_FT} we show the corresponding Fourier spectra of the occupation dynamics $\langle \hat{n}_l(t) \rangle$ in \fref{app:fig:occs_uniform_FT}.
All oscillation modes contributing to the time evolution of $\langle \hat{n}_l(t) \rangle$ 
show up as sharp peaks that strongly dominate an almost vanishing background noise in the case of a quantum wave packet that is launched on the staggered dimer configuration.
This has to be contrasted to the more uniform and shallow Fourier spectra that are found when Fock states are time evolved (shown in black).
We compare the same four Fourier spectra of increasing $\gamt$, both with time scaled with $\lambda_{\mathrm{max}}(\gamt)$ (left panels) and with $T_{\rm cl}(\gamt)$ (right panels).
In the regime of very weak interactions $\gamt \lesssim 0.2$ (upper four panels), all dominant peaks scale almost identically with $\gamt$ (and thus also with $1/\lambda_{\mathrm{max}}$ and $T_{\mathrm{cl}}$).
A closer look on the Fourier spectra for increasing $\gamt$ reveals that the scaling of the dominant frequencies lies somewhere between the $T_{\rm cl}$ scaling and the $1/\lambda_j$ scaling, but with varying weights and thus breaking uniformity.
This strongly indicates the existence of various classical periodic orbits that oscillate between the vicinities of the two symmetry-related staggered-dimer orbits, as their periods would be strongly influenced by both the period $T_{\rm cl}$ of the approached staggered-dimer orbits and the time scale $1 / \lamL$ for leaving (and approaching) the vicinity of the latter.

\sectionQ{Scars versus tunneling oscillations between stable regions: A comparison to the self-trapped regime}

\begin{figure}
  \centering
  \includegraphics[width=0.9\linewidth]{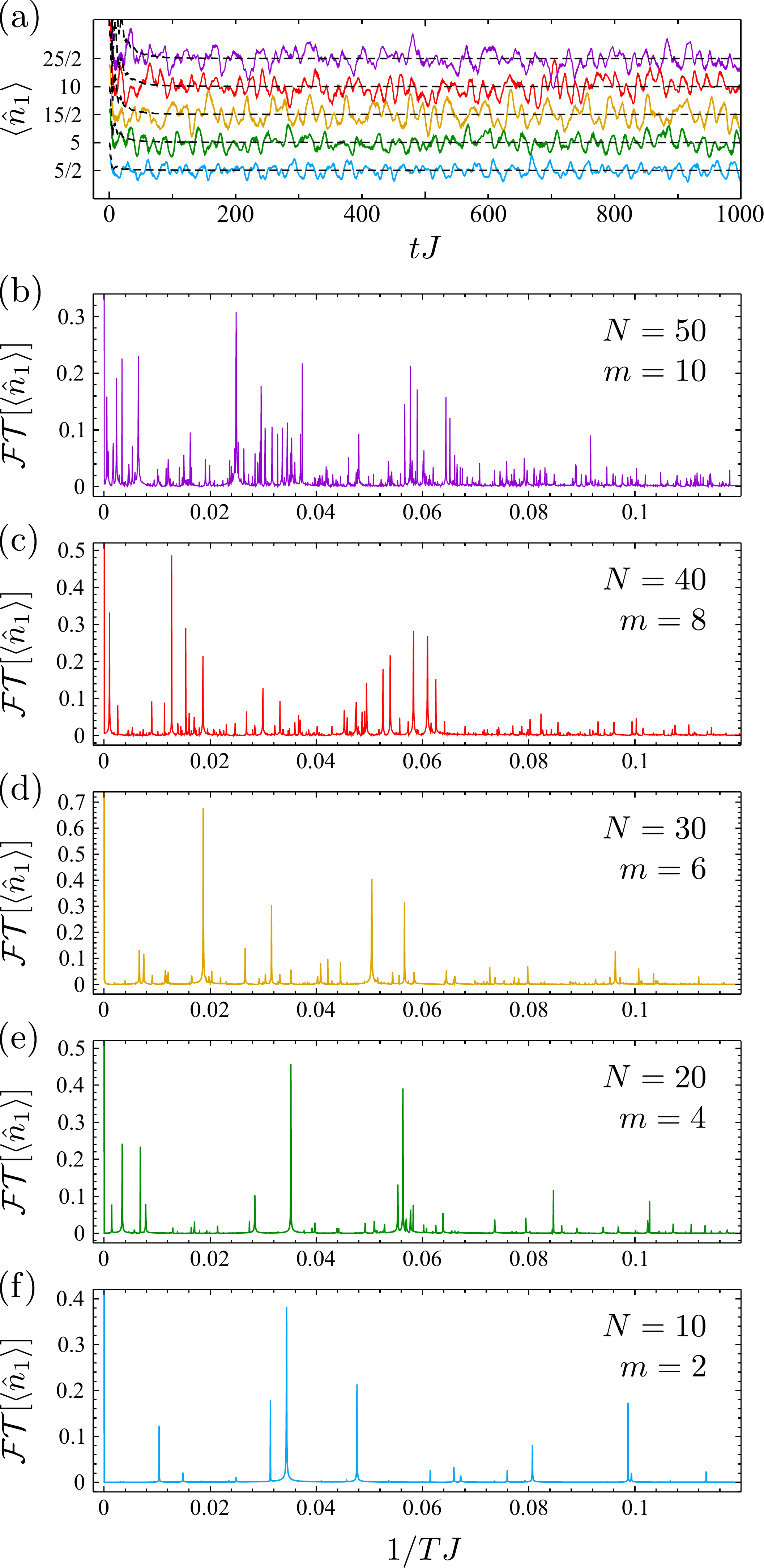}
  \caption{\label{app:fig:occs_Ndep}
    (a) Persistent oscillations in the single-site occupancy $\langle \hat{n}_1(t) \rangle$ of staggered-dimer product states $\lvert \pi_m \rangle$ [see \Leref{eq:prod}] for various total particle numbers $N=10,20,30,40,50$ (bottom to top) with fixed scaled interaction parameter $\gamt = 0.8$.
    The index $m$ of the product states is scaled such that the corresponding classical staggered dimer orbit is situated at a fixed imbalance of $z=0.6$.
    The oscillations about the thermal average do not decay over time and show an amplitude that does not vanish for inrcreasing $N$. This suggests that observability is maintained for large $N$ despite the increasing number of contributing scarred eigenstates.
    In contrast, the truncated Wigner approximation (black dashed) undergoes fast thermalization.
    (b)--(f); Corresponding discrete Fourrier transforms of $\langle \hat{n}_1(t) \rangle$ for $N=10$--$50$ show that the dominant frequencies of the oscillations stay in a band that does not change strongly with $N$.
  }
\end{figure}

In Appendix \ref{app:sec:uniform} we have identified strong indicators for the time scales of the persistent oscillations to be traced back to underlying classical quantities.
To further substantiate this, in \fref{app:fig:occs_Ndep} we investigate their dependence on the total particle number $N$.
By fixing $\gamt = 0.8$ and the imbalance $z=0.6$ of the initial staggered dimer state, $N$ serves as an effective inverse quantum of action (effective Planck's constant), while the purely classical dynamics is independent of $N$ (up to scaling).
Looking at the Fourier spectra of different $N$ (b)--(f), a direct comparison of individual peaks such as in \fref{app:fig:occs_uniform_FT} is not possible here, because the graining of the spectrum of the Hamiltonian and thus the available frequencies depend strongly on the Hilbert space dimension.
However, we find that the dominant frequencies stay in bands that are not varying strongly with $N$.
This clearly discriminates the observed oscillations from quantum phenomena such as tunneling, where time scales depend \emph{exponentially} on the inverse quantum of action.

Moreover, we find that the absolute amplitude of the oscillations around the ergodic average apparently does not decrease with increasing $N$, as can be seen in panel (a), although more and more frequencies get involved.
In a semiclassical picture, more and more ``mediating orbits'' that oscillate between the two symmetry-related staggered-dimer orbits are fulfilling the BS quantization condition.
Since their scaled classical actions $S/N$ (which is a classical invariant under fixing $z$ and $\gamt$) is quantized in steps of $\mathcal{O}(1/N)$, the number of contributing quantized orbits can roughly be estimated to scale linearly with $N$, while the local Hilbert space dimension scales $\sim N^3$.

In the following, we further contrast persistent oscillations originating from genuine quantum scarring as observed here from tunneling oscillations between classically regular islands as in a double-well scenario.
For an example of the latter, we tune the Bose-Hubbard plaquette into the self-trapping parameter regime. 
To be precise, we fix $\gamt = 9.5$ and initialize the system in imbalanced Fock states
\begin{equation}
    \lvert \phi( t = 0 ) \rangle = \lvert N/2, 0, N/2, 0 \rangle
\end{equation}
with alternating occupied and unoccupied sites, probing different (even) particle numbers $N$.
In contrast to the staggered-dimer states $\lvert \pi_m \rangle$, these do not exhibit a fixed phase relation between opposite sites.
We choose these states reflecting the fact that we are in the regime of Mott insulation.

For a better understanding, consider the corresponding two-site Bose-Hubbard system, in which self-trapping starts at $\gamt = 2$ with a bifurcation in the classical phase space:
The stable fixed point at $\bpsi = ( \sqrt{N/2}, -\sqrt{N/2} )$ becomes unstable and two symmetric stable fixed points appear, the points of classical self-trapping.
With growing $\gamt$ they depart towards the ``north-'' and ``south pole'' of the Bloch sphere, thereby enclosed by an ever growing separatrix.
The latter separates two symmetric islands of elliptic motion around the stable fixed points and touches the north pole $\bpsi = ( \sqrt{N}, 0 )$ and the south pole $\bpsi = ( 0, \sqrt{N} )$ at $\gamt = 4$.
For $\gamt = 9.5$, it has grown well beyond the poles, such that the latter are now embedded deeply in the stable islands of elliptic motion.
Points initially around the poles move along deformed circular orbits.
For $\gamt \to \infty$, these would become perfectly horizontal circular orbits, corresponding to Fock states as their tube-state counterpart.
For finite but strong interaction, $\gamt = 9.5$, they are deformed and tilted, but only slightly varying in the imbalance $z$ corresponding to the latitude on the Bloch sphere.

\begin{figure}
  \includegraphics[width=0.98\linewidth]{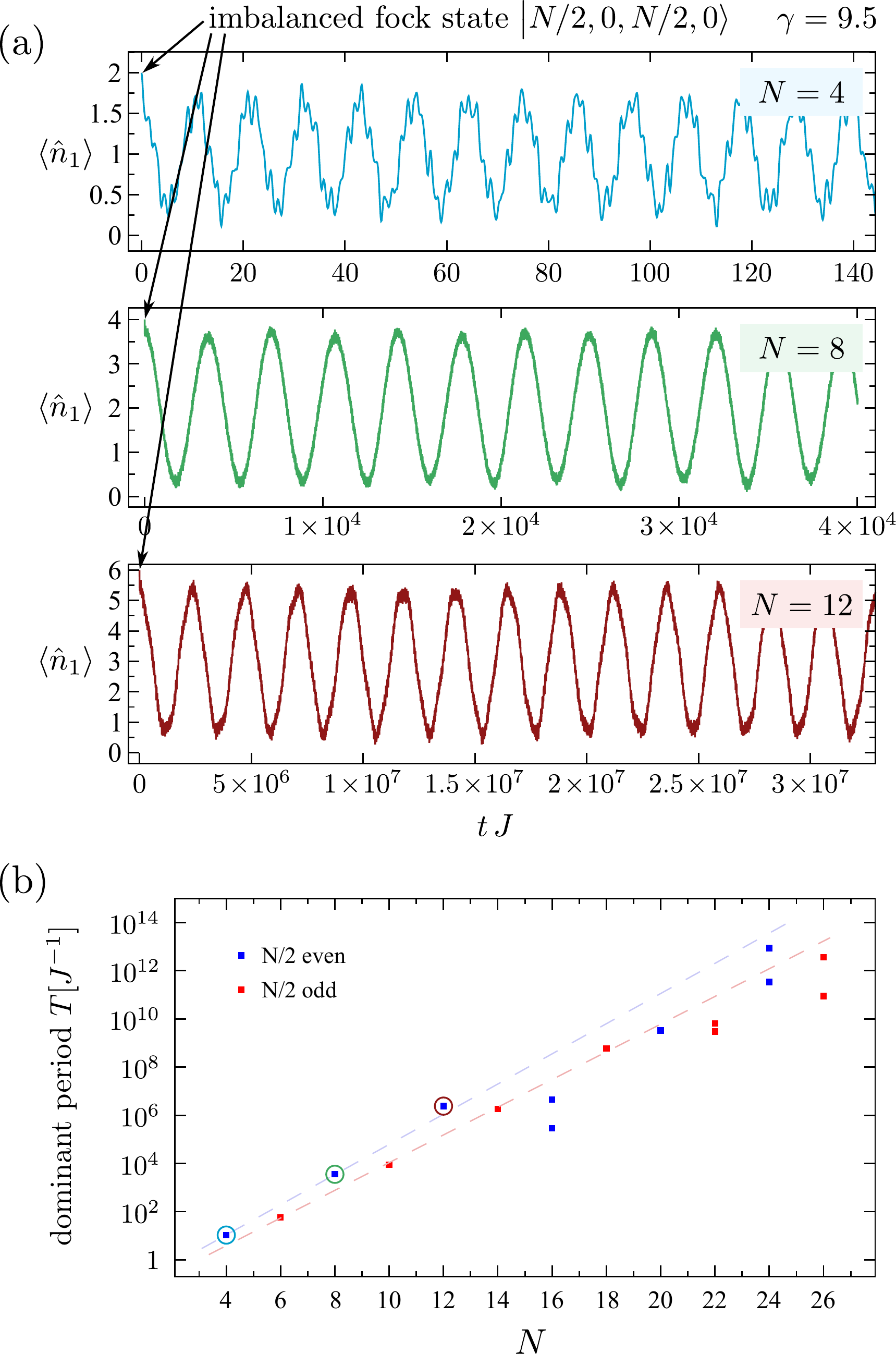}
  \caption{\label{app:fig:occ_STtunneling}
    (a) Oscillations of the expectation value of a single-site occupancy in the strongly interacting ($\gamt = 9.5$) Bose-Hubbard plaquette ($L=4$).
    The system is initialized in the Fock state $ \lvert \phi \rangle = \lvert N/2, 0, N/2, 0 \rangle$ for varying particle numbers $N = 4, 8, 12$.
    The distinct regularity and strongly varying periods clearly indicate that the observed oscillations originate from tunneling between symmetrically separated regular islands (see text).
    (b) The dominant periods of these oscillations, extracted for various $N$ (blue: even, red: odd), match the typical exponential decrease of tunneling rates with the quantum of action (see text).
    The blue and red dashed lines correspond to linear exponentials to guide the eye.
  }
\end{figure}

The two-site phase space is fully embedded in the four-site system as the low-dimensional manifold of two-site symmetric (or dimer-) states of the form $\bpsi = ( \psi_1, \psi_2, \psi_1, \psi_2 )$.
While further apart from this manifold, non-trivial four-site dynamics takes place, the dynamics within this manifold is exactly given by the two-site dynamics.
This implies that the self-trapped fixed points of the two-site system have fixed points as their counterparts in the four-site system, or any Bose-Hubbard ring of even sites $L \in 2 \mathbb{N}$ by periodic repetition.
Moreover, the stability of these fixed points is also partially inherited from the elliptic motion of the two-site system.
For $L=4$ and $\gamt=9.5$ the two self-trapping fixed points, located at $\psi_{1,2} = \pm \sqrt{(N/L+1/2)(1 \pm z)}$ with imbalance $z = z_0 \equiv \sqrt{1-4/\gamt^2}$ and $z = - z_0$, have six fully vanishing stability exponents.
This leaves only a single stable/unstable pair with minor instability.
Our simulations suggest that the latter originates from a residual non-trivial coupling of the relative phases.
Only for extremely strong interactions the phases are fully decoupled, turning the remaining stable/unstable dynamics into meta-stable behavior.
One signature of this is that the corresponding deviation $\delta \mathbf{z}(t)$ [see Eq.~\eref{app:eq:deltaz}], or more precisely the corresponding singular value of $\mathbb{M}_t$, is growing only linearly in time instead of exponentially.
Here, we see apparently a precursor of this meta-stability in the form of two disconnected thin locally chaotic patches around the two fixed points, stretched in the direction of altering relative phases.

In \fref{app:fig:occ_STtunneling}a we show the resulting occupation oscillations in the four-site plaquette for particle numbers $N = 4, 8$ and $12$.
Note the distinct regularity and strongly differing time scales.
This regularity manifests itself in very few (mostly one or two) strongly dominating frequencies in the corresponding Fourier spectra.
We have extracted the most dominant frequencies for a wide range of particle numbers.
The corresponding periods are displayed on a logarithmic scale in \fref{app:fig:occ_STtunneling}b, implying an exponential dependence on the inverse effective Planck constant, $N$.
The dashed lines correspond to linear exponentials to guide the eye.
The outliers with enhanced rate are reminiscent of resonance phenomena in quantum tunneling.

To conclude, the simulations strongly indicate that regular occupancy oscillations as observed in \fref{app:fig:occ_STtunneling} are to be attributed to tunneling oscillations between two classically separated close-to-regular islands. They are here symmetrically arranged about the points $\bpsi = ( \sqrt{N/2}, 0, \sqrt{N/2}, 0 )$ and $\bpsi = ( 0, \sqrt{N/2}, 0, \sqrt{N/2} )$, or correspondingly the tori given by
\begin{equation}
    \bpsi = ( \sqrt{N/2}\rme^{\rmi \theta_1}, 0, \sqrt{N/2}\rme^{\rmi \theta_2}, 0 )
\end{equation}
and
\begin{equation}
    \bpsi = ( 0, \sqrt{N/2}\rme^{\rmi \theta_1}, 0, \sqrt{N/2}\rme^{\rmi \theta_2} )
\end{equation}
with $\theta_{1,2}$ running through $(0,2\pi]$.
They are clearly to be distinguished from the genuine quantum scarring we have found for staggered dimer states.

\sectionQ{Scarring in larger systems}

\subsection{Periodic repetition of classical dynamics --- quantum scarring for $L=8,12$}

Regarding the classical dynamics, staggered dimer solutions are easily generalized from $L=4$ to $L= 4 \nu$ with integer $\nu$ by periodic extension $\psi_{l+4}(t) = \psi_l(t)$.
In order to investigate if correspondingly scarred eigenstates exist also in larger systems $\nu>1$, one cannot rely on a direct analysis of eigenstates.
Since the Hilbert space dimension grows exponentially with $L$ (keeping $N/L$ fixed), exact diagonalization becomes numerically intractable very soon.
However, investigating the time evolution of occupancy expectation values as a signature offers a means to numerically identify the existence of scarring in somewhat larger systems.
We realize this within a Krylov-type scheme that operates on a low-dimensional subspace at every finite but small time step and obtain extremely well-converged results (relative error $\sim 10^{-12}$) by increasing the dimension of the Krylov subspace or decreasing the time step.

\begin{figure}
  \centering
  \includegraphics[width=0.98\linewidth]{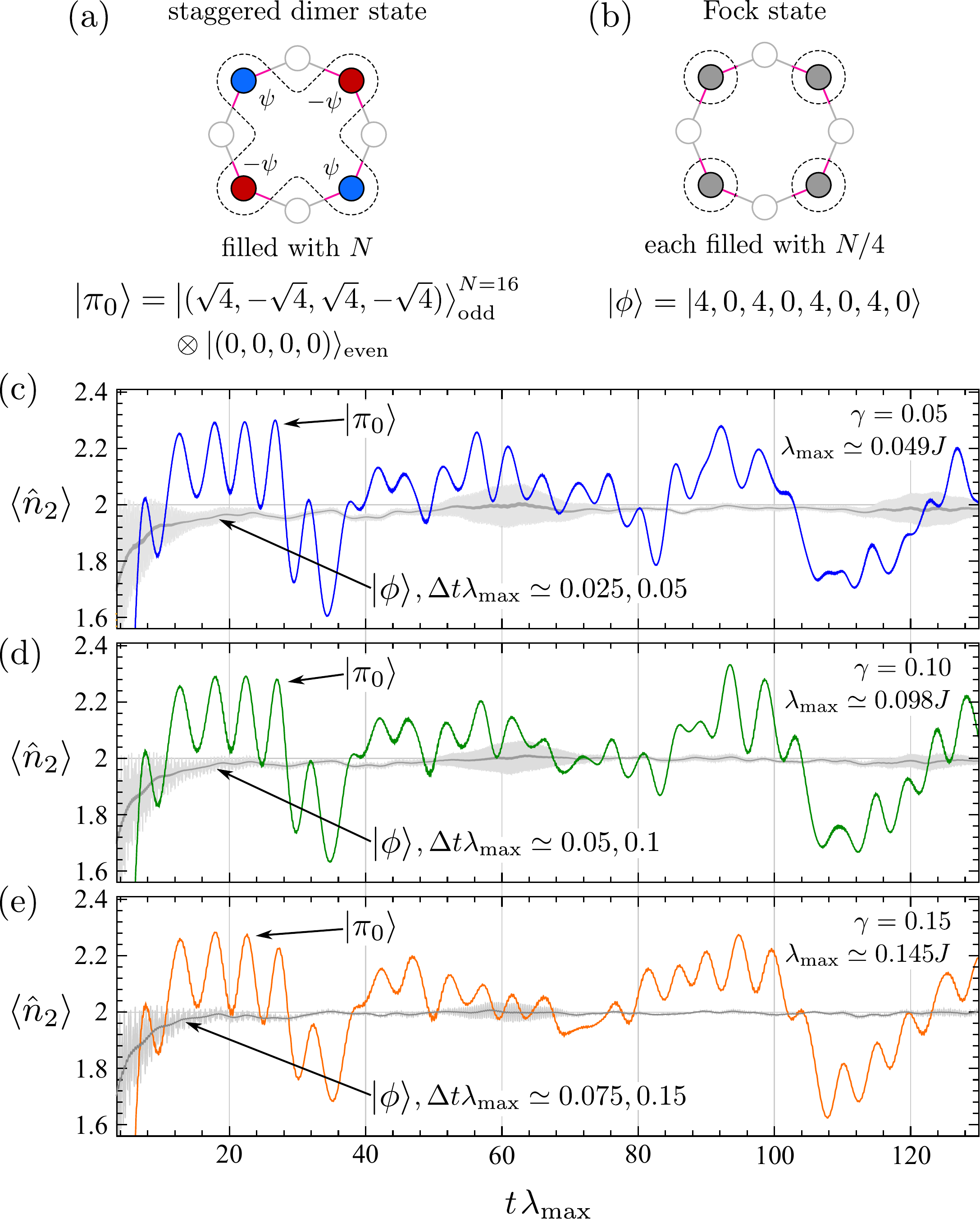}
  \caption{\label{app:fig:occs_L8}%
    Occupancy oscillations of $N=16$ atoms in the periodic Bose-Hubbard chain with $L=8$ sites in the regime of very weak interaction $\gamt=0.05, 0.10, 0.15$ (c)--(e).
    When preparing the system in staggered dimer states (a), scarring leads to persistent oscillations [colored in (c)--(e)].
    In contrast, an initialization in Fock states (b) shows fast thermalization (up to minor quantum revivals due to the finite dimensional Hilbert space).
    The data of the latter has been smoothed with a Gaussian of small width $\Delta t = 0.5 J^{-1}$ and $1.0 J^{-1}$ (light and dark gray, respectively) to suppress the fast quantum fluctuations on the scale of the hopping time $J^{-1}$, not present in the evolution of staggered dimer states due to the inherent suppression of hopping.
    As in the four-site case (see \fref{app:fig:occs_uniform}), a uniform time scale emerges which leads to a universal limiting wave form (see text).
  }
\end{figure}

In \fref{app:fig:occs_L8} we show that staggered dimer states in an eight-site chain filled with $N=16$ bosons exhibit non-thermalizing persistent oscillations very similar to the case $L=4$ (c.f., \fref{app:fig:occs_uniform} and \Lfref{fig:osc}).
As in the four-site system, we find that all positive stability exponents $\lambda^{\nu=2}_j$ (here $j=1,\ldots,6$) become equal in the regime of weak on-site interaction $\gamt \to 0$.
Moreover, two of them, say $\lambda_1$ and $\lambda_2$, (as functions of $\gamt$ and $z$) conincide exactly with those of the four-site system.
This exact inheritance originates in an embedding of the $L=4$ phase space as a submanifold within the full $L=8$ phase space by periodification:
Any trajectory $\bpsi^{L=4}(t) = (\psi_1(t), \ldots, \psi_4(t))$ that solves the DNLSE [see \Leref{eq:DNLSE}] of the four-site ring with a given interaction parameter $\gamt$ gives rise to a solution $\bpsi^{L=8}(t) = (\psi_1(t), \ldots, \psi_4(t), \psi_1(t), \ldots, \psi_4(t))$ in the corresponding eight-site ring with equal $\gamt$.
In particular, \emph{deviating} trajectories defining the stability spectrum about a given ``central'' orbit in $L=4$ are passed on to corresponding deviating trajectories (and thus stability exponents) in $L=8$ with four-site periodicity.
Analogously, all $16$ stability exponents of the eight-site ring are inherited by the sixteen-site ring and so on.
In addition, we conjecture that all other stability exponents not inherited by the smaller system assume the same equal values in the limit $\gamt \to 0$, as we have confirmed for small $\nu$.
For the stability spectrum of staggered dimer orbits this implies that all $\lambda_j \sim \gamt$ as $\gamt \to 0$, resulting again in the emergence of a uniform classical time scale, since the classical period $T_{\mathrm{cl}} = \pi/\gamma z J$ as inherited by the $L=4$ system has the same behavior.
This uniformity, or, more precisely, the resulting universal limiting wave form for occupancy oscillations, can clearly be seen by direct comparison of the quench dynamics for $\gamt=0.05, 0.10$, and $0.15$ in the $L=8$ ring.

The inheritance of stability spectra and the conjectured homogeneity has a strong impact on quantum scarring along staggered dimers in bigger rings.
For $\gamt \to 0$ the decrease of all (positive) stability exponents $\lambda_j \sim \gamt$ is exactly compensated by the increasing classical period $T_{\mathrm{cl}} \sim 1/\gamt$, such that the inverse multidimensional Heller indicator becomes
\begin{equation} \label{app:eq:HellerSD}
    \lamSpos T_{\mathrm{cl}} = \sum_{j=1}^{4 \nu - 2} \lambda_j T_{\mathrm{cl}} \to 2 \pi (2 \nu - 1) f(z) \,,
\end{equation}
with a universal function $f(z)$ that tends to $\simeq 1$ for maximum imbalance $z = 1$.
From a semiclassical perspective based on the Heller-type criterion (\ref{eq:Heller}),
$\lamSpos T_{\mathrm{cl}} \lesssim 2 \pi$, this has two consequences.
First, arbitrarily lowering the interaction will, most likely, not enhance quantum scarring beyond a certain point.
This is confirmed by the observation that the ``magnitude'' of scarring, as measured by the \emph{amplitude} of occupancy oscillations, reaches a stable plateau for weak interaction as seen in \fref{app:fig:occs_L8}.
Second, for increasing system size $\nu$, the increasing number of unstable directions makes quantum scarring less and less likely.
For $\nu=2$, i.e., $L=8$, we still find strong evidence for staggered dimer quantum scarring.
But the overall amplitude of occupancy oscillations is already diminished as compared to the four-site case.

\begin{figure}
  \centering
  \includegraphics[width=0.98\linewidth]{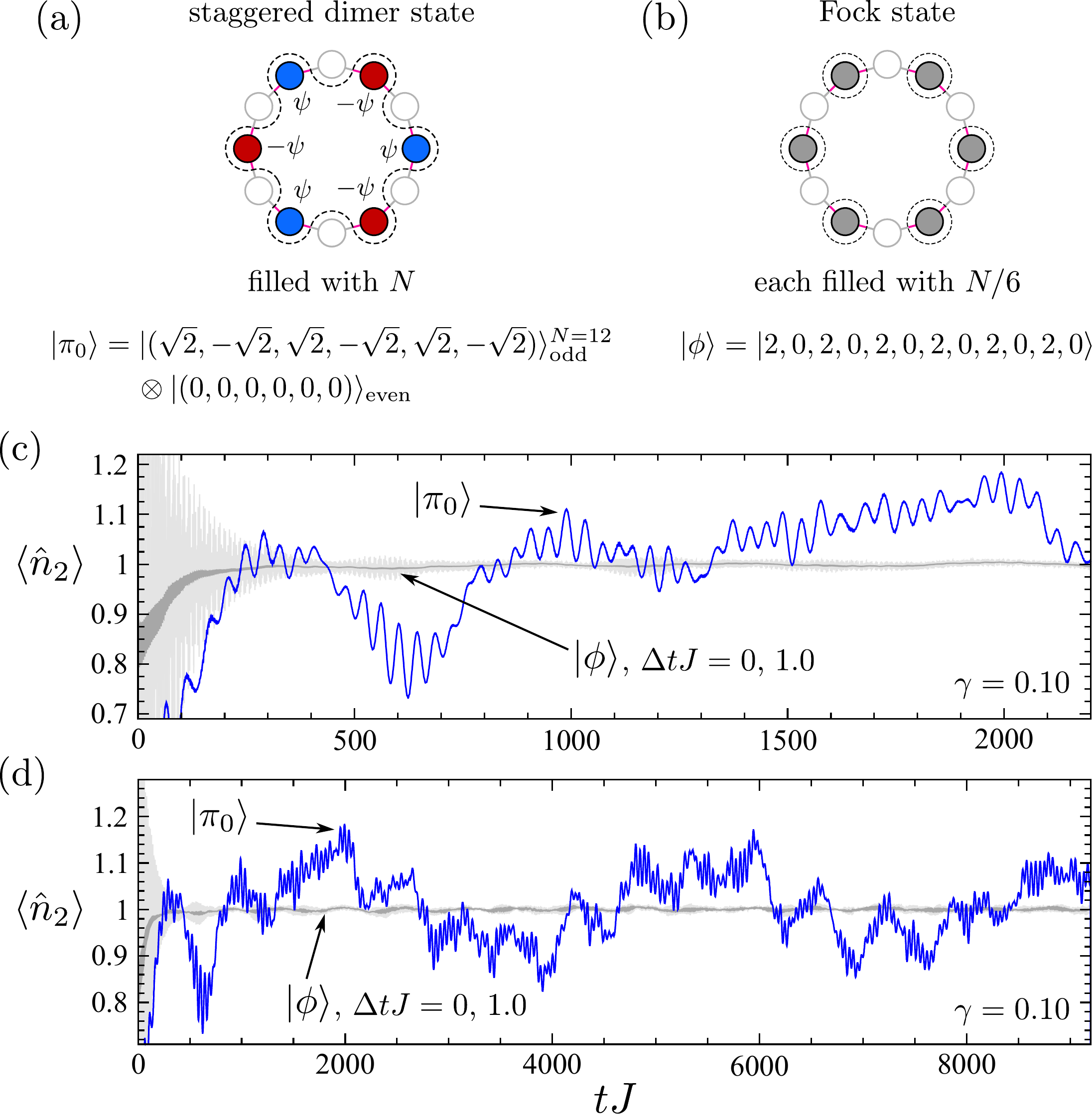}
  \caption{\label{app:fig:occs_L12}
    Occupancy oscillations of $N=12$ atoms in the periodic Bose-Hubbard chain with $L=12$ sites in the regime of very weak interaction $\gamt=0.10$ [(c) and (d)].
    When preparing the system in a staggered dimer state (a), scarring leads to oscillations [blue in (c) and (d)] which persist for long times (d).
    In contrast, an initialization in a corresponding Fock state (b) that lives in the same Hilbert subspace shows fast thermalization.
    The data of the latter are shown both raw and smoothed with a Gaussian of small width $\Delta t = J^{-1}$ (light and dark gray, respectively).
  }
\end{figure}

    As demonstrated in \fref{app:fig:occs_L12}, even in the corresponding $L=12$ site system with $N=12$ the occupancy oscillations traced back to scarring were still found to be present, with a magnitude of about $10\%$--$20\%$ with respect to the thermal average.
    In contrast, the latter is reached very fast with hardly visible remaining fluctuations in the case of an initial Fock state $\lvert \phi \rangle$.
    This is the expected behavior of a generic state in a high-dimensional Hilbert space.
    Both states live in the same invariant subspace $\mathcal{H}_0 \oplus \mathcal{H}_2$, which (at unit filling $N=L$) has a dimension of $\dim(\mathcal{H}_0) + \dim(\mathcal{H}_2) \simeq 10^5$ (see section \ref{app:sec:BHsym} below).
    On the one hand, the overall slightly lower magnitude of oscillations as compared to $L=8$ (with equal filling $N/L$) is in line with the smaller Heller-type indicator.
    On the other hand, it is surprising that scarring takes place at all for a Heller indicator as small as $2 \pi / \lamSpos T_{\mathrm{cl}} \simeq \frac{1}{5} \ll 1$.
    This apparent contradiction is resolved when accounting for the present discrete symmetries in the Heller criterion (see section \ref{app:sec:HCsym}).

\subsection{Symmetry-enhancement of quantum scarring --- thermodynamic limit}
\label{app:sec:BHsym}
    
    The Hamiltonian of the $L$-site Bose-Hubbard ring comes naturally in blocks that correspond to the irreducible representations of the underlying symmetry group, in this case the dihedral group $D_{2L}$ (we take $L$ even).
    It is generated by two elements, which can be taken as the rotation $r$ by one site and the reflection (inversion) $s$ on the first site.
    we will refer to the corresponding group actions on Hilbert space as $\hat{R} = \hat{U}(r)$ and $\hat{S} = \hat{U}(s)$.
    The total number of group elements is $\lvert D_{2L} \rvert = 2 L$.
    Our scar candidates, i.e., the classically periodic staggered dimer orbits, lie within a particular symmetry subspace of classical phase space, and thus the corresponding tube state, as elements of the Hilbert space, inherit specific symmetry properties. 
    As a consequence, the quantum dynamics of the tube states is generically restricted to an invariant subspace of the Hilbert space, given by one or the direct sum of several irreducible representations, or their components in case of irreps of dimension $s_\alpha > 1$.
    This restriction effectively lowers the available Hilbert space dimension and, as we have found in appendix \ref{app:sec:HCsym}, has a non-negligible enhancing effect on the scarring-likelihood.

    In the dihedral group with even number of sites $L$, the number of irreducible representations is $N_{\mathrm{ir}} = L/2 + 3$, which come as the four one-dimensional irreps $M^{(\alpha)}$, $\alpha = 0,1,2,3$ with
    \begin{equation}
        \begin{array}{ll}
            M^{(0)}( r ) = 1, \quad& M^{(0)}( s ) = 1, \\
            M^{(1)}( r ) = 1, \quad& M^{(1)}( s ) = -1, \\
            M^{(2)}( r ) = -1, \quad& M^{(2)}( s ) = 1, \\
            M^{(3)}( r ) = -1, \quad& M^{(3)}( s ) = -1,
        \end{array}
    \end{equation}
    and $L/2 - 1$ two-dimensional ones.
    The total dimension, i.e., the total number of independent symmetry subspaces $\mathcal{H}_{\alpha,i}$ of Hilbert space is thus $N_s = L + 2$.
    We denote with $\mathcal{H}_{0,1,2,3}$ the subspaces corresponding to the one-dimensional irreps $M^{(0)}, \ldots, M^{(3)}$.
    
    The staggered dimer product states $\lvert \pi_m \rangle$ that we use in the quench scenarios [see \Leref{eq:prod}] can be written as superpositions
    \begin{equation}
        \lvert \pi_m \rangle = \frac{1}{2} \left( \lvert \pi_m^+ \rangle \lvert + \lvert \pi_m^- \rangle \right)
    \end{equation}
    of the (non-normalized) states
    \begin{equation}
        \lvert \pi_m^\pm \rangle \equiv \lvert \pi_m \rangle \pm \lvert \pi_{N-m} \rangle \,.
    \end{equation}
    Throughout, we assume that the total particle number $N$ is even.
    The above states then have symmetry
    \begin{equation}
        \begin{split}
            \lvert \pi_m^+ \rangle \in \left\{ \begin{array}{lll}
                \mathcal{H}_0 \;\; &\text{for}\;\; &m \text{ even}\,, \\
                \mathcal{H}_3 \;\; &\text{for}\;\; &m \text{ odd}\,, \\
            \end{array}\right. \\
            \lvert \pi_m^- \rangle \in \left\{ \begin{array}{lll}
                \mathcal{H}_2 \;\; &\text{for}\;\; &m \text{ even}\,, \\
                \mathcal{H}_1 \;\; &\text{for}\;\; &m \text{ odd}\,, \\
            \end{array}\right.
        \end{split}
    \end{equation}
    implying for example that the maximally imbalanced state $\lvert \pi_0 \rangle$ lives in $\mathcal{H}_0 \oplus \mathcal{H}_2$ (and will stay in that subspace after the quench).
    At the same time, this is the subspace to which the dynamics of the corresponding Fock state $\lvert \phi \rangle$ (as used in the quench simulations) is confined.
    
    To find the symmetry enhancement factor $\gsym$ of the Heller-type indicator~\eref{app:eq:HellerSym} that should be applied to estimate the scarring likelihood of staggered dimer-like eigenstates, one needs to identify the symmetries of a single wave packet placed on the staggered dimer orbit, i.e., the number-projected coherent state $\lvert \bpsi \rangle_N$ with staggered dimer configuration $\bpsi = ( \psi_1, \psi_2, -\psi_1, -\psi_2, \ldots )$.
    This symmetry is different for $m=0,N$, i.e., $\psi_2 = 0$ or $\psi_1 = 0$, and $m \neq 0,N$.
    In the first case one can write $\lvert \bpsi \rangle_N$ as superposition of $\lvert \bpsi \rangle_N + \hat{R} \lvert \bpsi \rangle_N$ and $\lvert \bpsi \rangle_N - \hat{R} \lvert \bpsi \rangle_N$, which have symmetry $\alpha = 0$ and $\alpha = 2$, respectively, meaning that $\lvert \bpsi \rangle_N$ lives in $\mathcal{H}_0 \oplus \mathcal{H}_2$.
    In the second case, $m \neq 0,N$, $\lvert \bpsi \rangle_N$ has to be written as the superposition of four states that have symmetry $\alpha = 0, 1, 2, 3$.
    These four states are in turn constructed as linear combinations of $\lvert \bpsi \rangle_N$, $\hat{R} \lvert \bpsi \rangle_N$, $\hat{S} \lvert \bpsi \rangle_N$, and $\hat{R} \hat{S} \lvert \bpsi \rangle_N$ with coefficients $(1,1,1,1)$, $(1,-1,1,-1)$, $(1,1,-1,-1)$, and $(1,-1,-1,1)$ for the symmetries $\alpha=0,1,2,3$, respectively.
    This means that for $m \neq 0,N$, $\lvert \bpsi \rangle_N$ lives in the larger subspace $\mathcal{H}_0 \oplus \mathcal{H}_1 \oplus \mathcal{H}_2 \oplus \mathcal{H}_3$, the reason being that then the staggered dimer configuration has a rotational direction and thus is not symmetric under reflection as is the case for $m=0,N$.
    This difference manifests itself in the generally more pronounced scarring of the maximally imbalanced $m=0,N$ staggered dimers as compared to the partially imbalanced ones $m \neq 0,N$ that has been observed throughout.
    
    In a first rough estimate we could assume that in the highly excited spectrum all symmetry classes have similar density of states, which would yield a symmetry factor of
    \begin{equation}
        \gsym^{(\text{equal})} = \left\{
        \begin{array}{lll}
            \displaystyle
            \frac{N_s}{2} = \frac{L}{2} + 1 \;\; &\text{for}\;\; &m=0,N \,, \\ \\
            \displaystyle
            \frac{N_s}{4} = \frac{L}{4} + \frac{1}{2} \;\; &\text{for}\;\; &m \neq 0,N \,. \\
        \end{array} \right.
    \end{equation}
    In both cases the linear increase with $L$ counterbalances the decrease of the Heller-type indicator $ 2 \pi / \lamSpos T_{\mathrm{cl}} \simeq 2 / ( L - 2 )$ in the weakly interacting regime with a symmetry enhanced indicator
    \begin{equation}
        \frac{2 \pi}{\lamSpos T_{\mathrm{cl}}} \gsym^{(\text{equal})} \simeq \left\{
        \begin{array}{lll}
            \displaystyle
            \frac{L+2}{L-2}\;\; &\text{for}\;\; &m=0,N \,, \\ \\
            \displaystyle
            \frac{L+2}{2(L-2)} \;\; &\text{for}\;\; &m \neq 0,N \,. \\
        \end{array} \right.
    \end{equation}
    Under these assumptions the enhanced indicator for the maximally imbalanced staggered dimer ($m=0,N$) would \emph{always} be greater than one and tend towards the marginal value of unity in the thermodynamic limit $L\to\infty$.
    In contrast, the partially imbalanced case would drop below the critical value of unity already for $L=8$ and tend towards $1/2$ in the thermodynamic limit.
    
    A more sophisticated estimate for the symmetry-reduced densities of states in relation to the full one is to relate the dimensions $\dim \mathcal{H}_{\alpha}$, $\alpha=0,1,2,3$ of the corresponding subspaces to the full Hilbert space dimension $\dim \mathcal{H}$.
    To do so, one can build the subspaces from the Fock basis by superimposing a Fock state $\lvert \phi \rangle$ with all symmetry-group transformed versions of it with coefficients corresponding to the one-dimensional representation $M^{(\alpha)}$.
    We define the symmetrized Fock state
    \begin{equation} \label{app:eq:Tphi}
        \hat{T}_\alpha \lvert \phi \rangle = \sum_{g \in G} M^{(\alpha)}( g^{-1} ) \hat{U}(g) \lvert \phi \rangle \,.
    \end{equation}
    It can be easily checked that this state has symmetry $\alpha$, i.e., $\hat{T}_\alpha \lvert \phi \rangle \in \mathcal{H}_\alpha$ for any $\lvert \phi \rangle$.
    In the thermodynamic limit, i.e., $L \to \infty$ while keeping $N/L$ fixed, almost all Fock states are fully asymmetric, i.e., the addends in the sum~\eref{app:eq:Tphi} can be generically considered as $2L$ mutually distinct states.
    Each of them, when exchanged with $\lvert \phi \rangle$, will produce the same symmetrized state (up to a scalar factor).
    The dimension $\dim \mathcal{H}_\alpha$ thus tends towards the total Hilbert space dimension divided by $2L$ due to overcounting for all four one-dimensional irreps $\alpha=0,1,2,3$.
    The symmetry enhancement of the Heller-type indicator from the analysis of subspace dimensions, valid for large $L$, can thus be estimated as
    \begin{equation}
        \gsym^{(\text{dim})} = \left\{
        \begin{array}{lll}
            \displaystyle
            \frac{\dim \mathcal{H}}{2 \dim \mathcal{H}_0} = L \;\; &\text{for}\;\; &m=0,N \,, \\ \\
            \displaystyle
            \frac{\dim \mathcal{H}}{4 \dim \mathcal{H}_0} = \frac{L}{2} \;\; &\text{for}\;\; &m \neq 0,N \,, \\
        \end{array} \right.
    \end{equation}
    with an even more optimistic view towards staggered-dimer quantum scarring in the thermodynamic limit, as the enhanced Heller indicator would tend to $2$ or $1$ for the maximally or partially imbalanced case, respectively.

\subsection{Dimer plane waves as additional candidates for scarring}

Moreover, the linear growth of the \emph{bare} Heller indicator~\eref{app:eq:HellerSD} for weakly interacting staggered dimer states with system size
is based on the exact compensation of the (single-channel) instability and the orbit period as $\gamt \to 0$, which is a peculiarity of staggered dimer orbits.
This also leaves open the possible existence of genuine quantum scars along other orbits in large BH chains.
For example, for even $L \ge 6$, the classical staggered dimer configurations can be generalized to \emph{dimer plane waves}, where adjacent dimers have a relative phase $\theta$ other than $\pi$.
To be precise, the corresponding classical periodic orbits are of the form
\begin{equation} \label{app:eq:dpw}
    \bpsi = ( \underbrace{\psi_1, \psi_2}_{\mathclap{\text{one dimer}}},
              \rme^{\rmi \theta} \psi_1, \rme^{\rmi \theta} \psi_2,
              \rme^{\rmi 2 \theta} \psi_1, \rme^{\rmi 2 \theta} \psi_2,
              \ldots )
\end{equation}
with a phase satisfying $\theta L / 2 = 0 \pmod{2\pi}$.
A trajectory initialized in this way remains of this form for all times, only changing in the single-dimer degrees of freedom $\psi_1$ and $\psi_2$.
The dynamics of $\psi_1(t)$ and $\psi_2(t)$ following from the full DNLSE applied to \eref{app:eq:dpw} is governed by effective two-site BH physics.
For convenience, we define $( \tilde{\psi}_1, \tilde{\psi}_2 ) \equiv ( \psi_1, \rme^{-\rmi \theta / 2} \psi_2 )$.
In these phase-shifted variables the equations of motion read
\begin{equation}
    \begin{split}
        \rmi \dot{\tilde{\psi}}_1 &= - 2 J \cos( \theta / 2 ) \tilde{\psi}_2  + U (\tilde{\psi}^\ast_1 \tilde{\psi}_1 - 1) \tilde{\psi}_1 \,, \\
        \rmi \dot{\tilde{\psi}}_2 &= - 2 J \cos( \theta / 2 ) \tilde{\psi}_1  + U (\tilde{\psi}^\ast_2 \tilde{\psi}_2 - 1) \tilde{\psi}_2 \,,
    \end{split}
\end{equation}
with only \emph{partially} suppressed hopping $J \cos(\theta/2)$ if $\theta \neq \pi$.
This means that in contrast to staggered dimer orbits, dimer plane waves do in general not feature a diverging orbit period $T_\mathrm{cl}$ when $\gamt \to 0$, while all stability exponents must continuously go to zero as one approaches the integrable limit of non-interacting atoms.
Thus, the multidimensional Heller criterion (\ref{eq:Heller}) could possibly be fulfilled in the thermodynamic limit of arbitrarily large systems as long as the interaction parameter $\gamt$ is chosen small enough, even without symmetry considerations.
Such dimer plane waves thus offer another intriguing possibility of finding genuine quantum scarring in large-scale many-body systems.
We leave the thorough investigation of these promising circumstances for future research.

\end{document}